\documentclass[10pt,a4paper]{article}
\usepackage{arxiv}
\usepackage[english]{babel}
\usepackage[utf8]{inputenc} % allow utf-8 input
\usepackage[T1]{fontenc}    % use 8-bit T1 fonts
\usepackage{url}            % simple URL typesetting
\usepackage{booktabs}       % professional-quality tables
\usepackage{nicefrac}       % compact symbols for 1/2, etc.
\usepackage{microtype}      % microtypography
\usepackage{lipsum}
\usepackage{color} % Can be removed after putting your text content
\usepackage{graphicx}
\usepackage{amsmath}
\usepackage{amsfonts}
\usepackage{amssymb}
\usepackage{mathptmx}
\usepackage{txfonts}
\usepackage{ae,aecompl}
\usepackage{placeins}
\usepackage{hyperref}

\hypersetup{colorlinks,
citecolor=blue
}
\usepackage[round]{natbib}
\usepackage{pdfpages}
\title{Stability of two-fluid galactic disc under the influence of an external tidal field.}

\author{
  K. Aditya\\
      Indian Institute of Science Education and Research, Tirupati 517507, India \\
  \texttt{E-mail : kaditya@students.iisertirupati.ac.in} \\
  }

\begin{document}
\maketitle

\begin{abstract}
We consider the dynamics of rotationally supported thin galactic disc composed of stars and gas under the influence of external 
tidal field and derive the coupled differential equations governing the evolution of instabilities. Further linearising the governing equation 
a modified dispersion relation and stability criterion for appraising the stability of the two fluid galactic disc under the influence of external tidal field 
is obtained. Possible applications and method for the same are discussed.
\end{abstract}


\section{Introduction}
Local stability of the galactic disc is a subtle balance between the mass contained in the fluid packet, the random velocity dispersion and the 
differential rotation. The competetion between the stabilising agents i.e differential rotation and the random velocity
dispersion and the destabilising agent i.e the mass content  is classically quantified by the stability criterion proposed by 
\cite{toomre1964gravitational}.
\begin{equation}
Q=\frac{\kappa \sigma}{\pi G \Sigma}
\end{equation}
Where $\rm \kappa$ is the epicyclic frequency, $\rm \sigma$ is the radial velocity dispersion, and $\rm \Sigma$ the surface mass density. Value of 
$\rm Q \geq 1$ quantifies the stability of galactic disc against axis symmetric perturbations. The the above criterion appraises the stability a fluid 
disc consisting of either gas or stars at a time, but a real galactic disc consists of stars and gas which is also gravitationally important, more so in 
case of gas rich low surface brightness galaxies (LSBs) \cite{de1996h}. The dispersion relation pertaining to the stability of gravitationally coupled two-fluid disc (stars+gas) has been 
extensively studied by \cite{jog1984galactic} and criterion for appraising the stability of two-fluid disc has been derived and studied by 
\cite{elmegreen1995effective}, \cite{jog1996local}, and the same for multi-component disc has been studied by \cite{rafikov2001local}. Stability 
criterion for multi-component galactic disc with finite thickness has been studied by \cite{romeo2013simple}, which is useful when the galactic disc 
consists of say mutiple stellar disc (thin disc+thick disc) and HI disc. The role of external tidal 
field on the stability of gravitating, infinite,homogenous systems was investigated by \cite{jog2013jeans} and modified Toomre stability criterion due to
external tidal field also derived by \cite{jog2014effective}. In this work we consider the simple case of an gravitationally interacting 
axis-symmetric isothermal fluid disc of stars and gas supported by differential rotation and the random velocity dispersion, under 
the influence of an external tidal field. We will derive the basic differential equations governing the growth of density perturbation 
and further obtain stability criterion for the two-component fluid disc in presence of external tidal field.
The paper is organised as follows: in section 2 we will formulate the basic equations and derive the governing differential equation, dispersion 
relation modified due to external tidal fields, and an effective stability parameter, in section 3 we will discuss possible applications and conclude 
in section 4.

\section{Formulation and derivation of basic equations}
The stars and gas in the galactic disc togather constitute isothermal fluids interacting with each other gravitationally. The fluid disc is supported 
by random presure and rotation. The problem is completely described by understanding the flow of perturbed fluid components constituting a thin 
cylindrical disc, i.e lying in z=0 plane. The dynamics of the perturbed fluid and the conservation of mass density are dictated by the linearised 
force equation (eq 2,3) and the linearised continuity equation (eq 4) for the two -fluid system in cylindrical coordinates. 
The presence of an external tidal field is included by adding an extra force $\frac{-\partial \phi_{ext}}{\partial r}$ in the linearised force 
equation. The dynamics of the fluid in force equation is governed by the self gravity of the two-fluid disc, i.e each component of the perturbed 
fluid say 'gas' moves under the combined potential of stars + gas, thus the right hand side of the linear force equation is supplimented with the 
linearised Poisson equation (eq 5) describing the combined mass densities of the two-fluid system and an isothermal 
equation of state $\rm P_{i} = \Sigma_{i} c_{i}^{2} $ is used to describe the presure  of the perturbed fluids, where $\Sigma_{i}$ denotes the 
surface density  and $\rm c_{i}$ denotes the isothermal speed of sound of the of the $i^{th}$ fluid component respectively. Throughout this work $\rm 'i'$ will index 
stars $(s)$ and gas $(g)$.

The linearised force equations under influence of an external tidal force are;
\begin{equation}
\frac{\partial u_{i}}{\partial t} + \Omega_{0}\frac{\partial u_{i}}{\partial \theta} - 
 2 \Omega_{0} v_{i} + \frac{c_{i}^{2}}{\Sigma_{0i}}\frac{\partial \Sigma_{i}}{\partial R} +\frac{\partial (\phi_{g} + \phi_{s})}{\partial R}
 +\frac{\Sigma_{i}}{\Sigma_{0i}} \frac{\partial \phi_{ext}}{\partial R} =0
\end{equation}

\begin{equation}
\frac{\partial v_{i}}{\partial t} -2Bu_{i} + \Omega_{0} \frac{\partial v_{i}}{\partial \theta} + 
\frac{1}{R} \frac{\partial (\Sigma_{s} + \Sigma_{g} )}{\partial \theta} = 0
\end{equation}

Linearised continuity equation reads;

\begin{equation}
\frac{\partial \Sigma_{i}}{\partial t} +\Omega_{0}\frac{\partial \Sigma_{i}}{\partial \theta} + \frac{\partial(\Sigma_{0i} u_{i})}{\partial R}
+ \frac{u_{i} \Sigma_{0i}}{R} +\frac{\Sigma_{0i}}{R} \frac{\partial v_{i}}{\partial\theta} =0
\end{equation}

Linearised Poisson equation reads;
\begin{equation}
\frac{1}{R}\frac{\partial}{\partial R} R \frac{\partial (\phi_{s} +\phi_{g})}{\partial R} + \frac{\partial^{2} (\phi_{s} + \phi_{g})}{\partial z^{2}} 
=4 \pi G( \Sigma_{s} + \Sigma_{g}) \delta(z)
\end{equation}

Here,$u_{i}$ and $v_{i}$ are the perturbed velocities in radial and tangential direction, $\Sigma_{i}$ and $\phi_{i}$ are the perturbed surface 
densities and the potentials of the $i^{th}$ components. The quantities $\Sigma_{0i}$ , $\Omega_{0}$ and $\phi_{0i}$ are the unperturbed surface 
densities, angular velocity and potential respectively.

Considering axis-symmetric case $\frac{\partial}{\partial{\theta}} \rightarrow 0 $, the perturbed equations read;

\begin{equation}
\frac{\partial u_{i}}{\partial t} -2 \Omega_{0} v_{i} + \frac{c_{i}^{2}}{\Sigma_{0i}} \frac{\partial \Sigma_{i}}{\partial R} + 
\frac{\partial (\phi_{s} +\phi_{g} )}{\partial R} + \frac{\Sigma_{i}}{\Sigma_{0i}} \frac{\partial \phi_{ext}}{\partial R} = 0
\end{equation}

\begin{equation}
\frac{\partial v_{i}}{\partial t} -2Bu_{i}=0
\end{equation}

\begin{equation}
\frac{\partial \Sigma_{i}}{\partial t} + \Sigma_{0i} \frac{\partial u_{i}}{\partial R}=0
\end{equation}
and the Poisson equation for the thin disc assumes the form \cite{1964ApJ...139.1217T};

\begin{equation}
\frac{\partial(\phi_{s} +\phi_{g})}{\partial R} = - 2 \pi i G(\Sigma_{s} + \Sigma_{g})
\end{equation}

Equations 6,7,8,9 will be basis for deriving the dynamic equations governing growth of perturbations and also the associated dispersion relation.
Indexing equation 6,7,8,9 for stars; 

\begin{equation}
\frac{\partial u_{s}}{\partial t} -2 \Omega_{0} v_{s} + \frac{c_{s}^{2}}{\Sigma_{0s}} \frac{\partial \Sigma_{s}}{\partial R} + 
\frac{\partial (\phi_{s} +\phi_{g} )}{\partial R} + \frac{\Sigma_{s}}{\Sigma_{0s}} \frac{\partial \phi_{ext}}{\partial R} = 0
\end{equation}

\begin{equation}
\frac{\partial v_{s}}{\partial t} -2Bu_{s}=0
\end{equation}

\begin{equation}
\frac{\partial \Sigma_{s}}{\partial t} + \Sigma_{0s} \frac{\partial u_{s}}{\partial R}=0
\end{equation}

%Operating with $\frac{\partial}{\partial R}$ on equation (10), and eliminating the terms 
%$\frac{\partial}{\partial R} \frac{\partial u_{s}}{\partial {t}}$ by taking time derivative of equation (12) which will give 
%$\frac{\partial}{\partial R} \frac{\partial u_{s}}{\partial {t}}= \frac{-1}{\Sigma_{0s}}\frac{\partial^{2} \Sigma_{s}}{\partial^{2}t}$. 
%And similarly $\frac{\partial v_{s}}{\partial R}$ is eliminating by operating $\frac{\partial}{\partial R}$ on equation (11) and 
%substituting for $\frac{\partial u_{s}}{\partial R}$ from equation (12) to get $\frac{\partial v_{s}}{\partial R} =-\frac{2B \Sigma_{s}}{\Sigma_{0}}$.
Combining the equations (9), (10), (11), and (12) we obtain the following equation for stars;
\begin{equation}
\frac{\partial^{2} \Sigma_{s}}{\partial t^{2}}  - c_{s}^{2} \frac{\partial^{2} \Sigma_{s}}{\partial R^{2}} -4\Omega_{0} B \Sigma_{s} + 
 2\pi i G \Sigma_{0s}\frac{\partial}{\partial R}(\Sigma_{s} + \Sigma_{g}) - \Sigma_{s} \frac{\partial^{2} \phi_{ext}}{\partial R^{2}}=0
\end{equation}
similar equation for gas reads ;
\begin{equation}
\frac{\partial^{2} \Sigma_{g}}{\partial t^{2}}  - c_{g}^{2} \frac{\partial^{2} \Sigma_{g}}{\partial R^{2}} -4\Omega_{0} B \Sigma_{g} + 
 2\pi i G \Sigma_{0g}\frac{\partial}{\partial R}(\Sigma_{s} + \Sigma_{g}) - \Sigma_{g} \frac{\partial^{2} \phi_{ext}}{\partial R^{2}}=0
\end{equation}

In order to obtain a dispersion relation we substitute  plane wave ansatz $e^{i(kr - \omega t)}$ for the  perturbed quantities in equations (13)
and (14), we obtain

\begin{equation}
\Sigma_{s} =\frac{2 \pi G k \Sigma_{0s} \Sigma_{g}}{(\omega^{2} -c_{s} ^{2} k^{2} -\kappa^{2} + 2 \pi G \Sigma_{0s} k - T_{0})}
\end{equation}
and similarly

\begin{equation}
\Sigma_{g} =\frac{2 \pi G k \Sigma_{0g} \Sigma_{s}}{(\omega^{2} -c_{g} ^{2} k^{2} -\kappa^{2} + 2 \pi G \Sigma_{0g} k - T_{0})}
\end{equation}

Where $T_{0}=-\frac{\partial^{2}\phi_{ext}}{\partial R^{2}} $. Combining equation (15) and (16) the final dispersion relation reads;

\begin{equation}
\begin{aligned}
(\omega^{2} -c_{s} ^{2} k^{2} -\kappa^{2} + 2 \pi G \Sigma_{0s} k - T_{0})
(\omega^{2} -c_{g} ^{2} k^{2} -\kappa^{2} + 2 \pi G \Sigma_{0g} k - T_{0})=\\
(2 \pi G \Sigma_{0s} k)(2 \pi G \Sigma_{0g} k)
\end{aligned}
\end{equation}

We now proceed to derive stability criterion a la \cite{jog1996local}. Defining the following parameters;
\begin{equation}
\begin{aligned}
\alpha_{s}= \kappa^{2}  + c_{s} ^{2} k^{2} -  2 \pi G \Sigma_{0s} k + T_{0}\\
\alpha_{g}= \kappa^{2}  + c_{g} ^{2} k^{2} -  2 \pi G \Sigma_{0} k + T_{0}\\
\beta_{s}=2 \pi G \Sigma_{0s} k\\
\beta_{g}=2 \pi G \Sigma_{0g} k
\end{aligned}
\end{equation}

Substituting above the dispersion relation and the respective roots are given by;
\begin{equation}
\begin{aligned}
\omega^{4} -\omega^{2}(\alpha_{s} + \alpha_{g})+(\alpha_{s} \alpha_{g} -\beta_{s} \beta_{g})=0\\
\omega^{2}_{\pm}=\frac{1}{2}(\alpha_{s} + \alpha_{g}) \pm \frac{1}{2}( (\alpha_{s}+\alpha_{g})^{2} -4(\alpha_{s} \alpha_{g} -\beta_{s} \beta_{g}))^\frac{1}{2}
\end{aligned}
\end{equation}
For a one component gaseous or stellar disc $\alpha_{g} \geq 0$ or $\alpha_{s} \geq 0 $ is sufficient condition for stability, whereas condition for
marginal stability of two-component disc reads $\omega^{2}=0$ or $(\alpha_{s} \alpha_{g} -\beta_{s} \beta_{g})=0$. And for the disc to be 
unstable the conditions is $\alpha_{s} \alpha_{g} -\beta_{s} \beta_{g} <0$.
With simple algebra the condition for neutral equilibrium can be written as;
\begin{equation}
\frac{2 \pi G \Sigma_{0s} k}{\kappa^{2} + k^{2}c_{s}^{2} +T_{0}} + \frac{2 \pi G \Sigma_{0g} k}{\kappa^{2} + k^{2}c_{g}^{2} +T_{0}} =F
\end{equation}
where F=1. For a marginally stable one fluid disc a function $\rm F$ can be defined as 
$\rm F= \frac{2 \pi G \Sigma_{0} k}{\kappa^{2} + k^{2}c^{2} +T_{0}}$, . The value of $k_{min}$ for the 1 fluid disc is obtained by putting $\frac{d\omega^{2}}{dk}=0$, where 
$\omega^{2}= \kappa^{2} - 2 \pi G \Sigma k +c^{2}k^{2} + T_{0}$, which yields $k_{min} = \frac{\pi G \Sigma}{c^{2}}$. Evaluating the 
polarisation function at $k_{min}$ yields $F= \frac{2}{1+Q^{2}}$.
In a manner similar to the one fluid case-fluid disc a function $\rm F$ for the marginally stable two-fluid 
disc is evaluated at $k_{min}$ which implies finding $\frac{d\omega^{2}_{-}}{dk}=0$,or $\frac{d(\omega^{2}_{+}\omega^{2}_{-})}{dk}=0$, 
i.e finding $\frac{d(\alpha_{s} \alpha_{g} -\beta_{s} \beta_{g})}{dk}$ which yields;
\begin{equation}
\begin{aligned}
k^{3}(4c_{s}^{2} c_{g}^{2}) - 3k^{2}(2\pi G\Sigma_{0s} c_{s}^{2} +2\pi G\Sigma_{0g} c_{g}^{2}) \\
+2k ((\kappa^{2}+T_{0})(c_{g}^{2}+c_{s}^{2}))
-(2\pi G\Sigma_{0s}+2\pi G\Sigma_{0sg}) (\kappa^{2}+T_{0})=0
\end{aligned}
\end{equation}
Now, defining $\epsilon = \frac{\Sigma_{0g}}{\Sigma_{0s}+\Sigma_{0sg}}$,and parameter $X=\frac{\kappa^{2}}{2 \pi G (\Sigma_{0s}+\Sigma_{0g})k_{min}}$, 
the criterion for appraising the stability of two-fluid disc under influence of external tidal field is given by;

\begin{equation}
\frac{2}{1+Q_{T}^{2}}=\frac{(1-\epsilon)}{X(1+ \frac{(1-\epsilon)^{2} Q_{s}^{2}}{4X^{2}} + \frac{T_{0}}{\kappa^{2}})} +
			\frac{\epsilon}{X(1+ \frac{\epsilon^{2} Q_{g}^{2}}{4X^{2}} + \frac{T_{0}}{\kappa^{2}})}
\end{equation}
The above equation can also be written as;

\begin{equation}
\frac{2}{1+Q_{T}^{2}}= \frac{(1-\epsilon)}{X'( 1+ \frac{(1-\epsilon)^{2} Q'^{2}_{s}}{4X'^{2}})}+
			\frac{\epsilon}{X'( 1+ \frac{\epsilon^{2} Q'^{2}_{g}}{4X'^{2}})}
\end{equation}
The primed quantities are defined as $X'= \frac{(\kappa^{2}+T_{0})}{2 \pi G (\Sigma_{0s}+\Sigma_{0g})k_{min}}$ and $Q'_{s}=\frac{\kappa c_{s}}{\pi G \Sigma_{0s}}(1+\frac{T_{0}}{\kappa^{2}})^\frac{1}{2}$ 
and  $Q'_{g}=\frac{\kappa c_{g}}{\pi G \Sigma_{0g}}(1+\frac{T_{0}}{\kappa^{2}})^\frac{1}{2}$. The original parameters describing the stability under influence 
of tidal field are modified as $X'=X(1+\frac{T_{0}}{\kappa^{2}})$, $Q'_{s}=Q_{s}(1+\frac{T_{0}}{\kappa^{2}})^\frac{1}{2}$ and $Q'_{g}=Q_{g}(1+\frac{T_{0}}{\kappa^{2}})^\frac{1}{2}$.
Where the stability parameter $Q_{T}$ is evaluated at $k_{min}$. $Q_{T}>1 ,=1, <1 $ correspond to stable, marginally stable and unstable
configurations of the two-fluid disc.
\section{Results/Discussion}
To gain better insight into on impact of tidal forces on two fluid disc we firstly investigate the impact of tidal force on a one-component disc.
The dispersion relation for a one-component disc in presence of tidal field is;
\begin{equation}
\omega^{2}=(\kappa^{2} + T_{0})k^{(0)} + c^{2}k^{(2)} -2 \pi G \Sigma k^{(1)}
\end{equation}
In the above equation at large value of $\rm k$,  $k^{2}$ will dominate thus presure stabilises the galactic disc at small scales, at small $\rm k$ 
i.e $\rm k^{(0)}$ the differential rotation and the tidal field stabilise the disc at large scales, and at intermdiate $\rm k$ self-gravity of the 
galactic disc becomes important. Tidal field comes in two flavours 'compressive' $\rm T_{0}<0$ and 'disruptive' $\rm T_{0}>0$. Classic example of disruptive 
tidal field  $\rm T_{0}>0$ is due to point mass. \cite{renaud2010dynamics}, appendix B, has an exhaustive compilation of tidal fields due to all the 
important potential-density pairs. A disruptive tidal field adds up with the differential rotation and presure to stabilise the disc, whereas a compressive tidal 
field adds up with the self gravity of the galactic disc and aids in destabilising the galactic disc. Next we proceed to inspect the marginal stability 
of the one-fluid galactic disc. Putting $\omega^{2} =0$ in equation (25) can be recast to obtain a quadratic equation in $\rm k$, 
\begin{equation}
1 + \frac{Q'^{2}}{4} \frac{k^{2}}{k'^{2}_{T}} -\frac{k}{k'_{T}}=0
\end{equation}
Where, $Q'=Q (1+\frac{T_{0}}{\kappa^{2}})^{\frac{1}{2}}$, $k'_{T}=k_{T}(1+\frac{T_{0}}{\kappa^{2}})$, $k_{T}=\frac{\kappa^{2}}{2  \pi G \Sigma_{0}}$ and 
defining $\zeta'= \frac{k'_{T}}{k}=\frac{k_{T}}{k}(1+\frac{T_{0}}{\kappa^{2}})$, i.e $\zeta'=\zeta(1+\frac{T_{0}}{\kappa^{2}})$.
With above substitutions equation (26) can be shown to assume;
\begin{equation}
Q=2\left[ \zeta \left( 1- \zeta (1+\frac{T_{0}}{\kappa^{2}})\right) \right]^{\frac{1}{2}} 
\end{equation}
\begin{figure}
\centering
\resizebox{65mm}{55mm}{\includegraphics{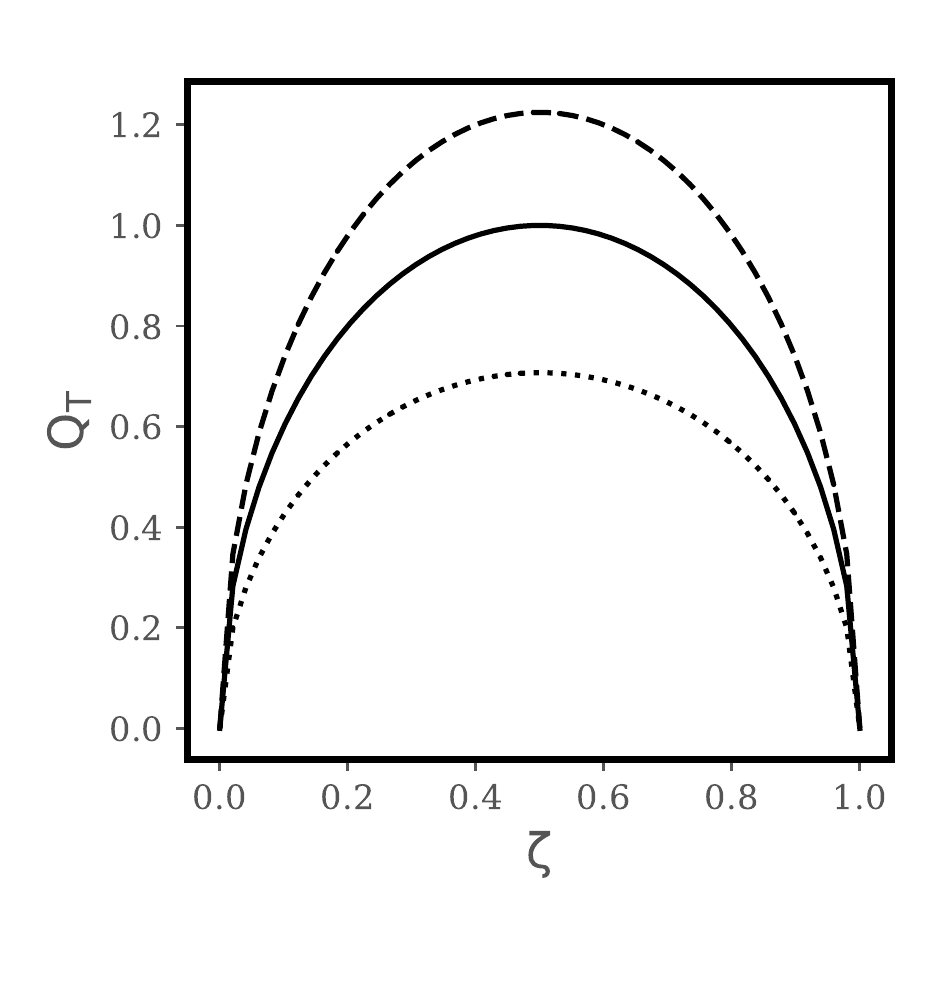}} 
\caption{The above figure depicts the marginal stability of the one fluid disc under influence of tidal field. 'Solid' line indicates 
marginal stabilty in absence of tidal field, 'dashed' line indicates marginal stabilty for disruptive tidal field $T_{0}>0$ and 'dotted' line indicates
stabilty of disc when acted upon by compressive tidal $T_{0}<0$.} 
\end{figure}

The marginal stability curve (Fig 1) is plotted fixing $\frac{T_{0}}{\kappa^{2}}$ at a fixed value of 0.5. It can be seen that for for value of $Q<1$
in absence of tidal field the disc becomes unstable, whereas disruptive tidal increase this value, so the disc is less succeptible to external perturbations.
And compressive tidal field lowers the value such that the disc becomes more prone to axis-symmetric perturbations.

Now we will try to address the role of compressive and disruptive tidal fields on a two fluid galaxy disc. We plot contours of $Q_{T}$, for the 
given values of $Q_{s}, Q_{g}, \epsilon,$ and $\frac{T_{0}}{\kappa^{2}}$. For the same range of values spanning 
$X_{s}=\frac{1}{2} (1-\epsilon) Q_{s}^{2}$ and $X'_{g}=\frac{1}{2} \epsilon Q_{g}^{2}$, and are computed 
over range of $Q_{s}$ and $Q_{g}$ at a constant value of $\epsilon$ and $\frac{T_{0}}{\kappa^{2}}$. $Q_{T}$ is evaluated by finding minimum
$X_{min}$ and maximum $X_{max}$  in range between $X_{s}$ and $X_{g}$, and finally $Q_{T}$ is evaluated over a range of $X_{max}$ and $X_{min}$, 
with the idea that $X$ lies in between $X_{max}$ and $X_{min}$. Procedure and applications of two-component stability formula without external 
tidal are discussed in greater detail in \cite{jog1996local}. For studying impact of tidal force we plot contour diagrams fixing gas-fraction
$\epsilon$  and $R=\frac{T_{0}}{\kappa^2}$ and for the same gas fraction we will vary $R$ to understand the impact of the external force on two-fluid 
disc.
In table (1) and table (2) are summarised the values of maximum and minimum values of $Q_{T}$ obtained for varying values of gas-fraction $\epsilon$  and strength 
of external tidal field.

\begin{table*}
\small
\centering
\begin{tabular}{|l|c|c|c|c|c|c|}
\hline
$R\rightarrow$ & 0.0 & +0.5  & +2.5 & -0.1 &-0.5  \\
$\epsilon \downarrow$ &         &     &    &   &\\
\hline    
\hline
0.05 &1.794 & 2.098 &2.693& -  &1.44\\
0.1 & 1.675 & 1.88&2.515&  -   &1.280\\
0.3& 1.306 & 1.599& 2.55&  1.88    &  -   \\
\hline
\end{tabular}
\caption{Above table describes the values of maximum values of $Q_{T}$ obtained for two fluid galactic 
disc under influence of external tidal fields by varying $T_{0}$ and $\epsilon$}
\label{table:A}
\end{table*}

\begin{table*}
\small
\centering
\begin{tabular}{|l|c|c|c|c|c|c|c|c|}
\hline
$R\rightarrow$ & 0.0 &+0.5 & +2.5 & -0.1&-0.5\\
$\epsilon \downarrow$ &         &    & &    &  \\
\hline    
\hline
0.05 &0.994 &1.198 & 1.693&  -  &0.647\\
0.1 &0.875&0.81&1.515&   -   &0.480\\
0.3&0.406&0.699&1.134& 0.388     &-\\
\hline
\end{tabular}
\caption{Above table describes the values of mimimum values of $Q_{T}$ obtained for two fluid galactic 
disc under influence of external tidal fields by varying $T_{0}$ and $\epsilon$}
\label{table:B}
\end{table*}

\begin{figure*}
\begin{center}
\begin{tabular}{cc}
\resizebox{65mm}{60mm}{\includegraphics{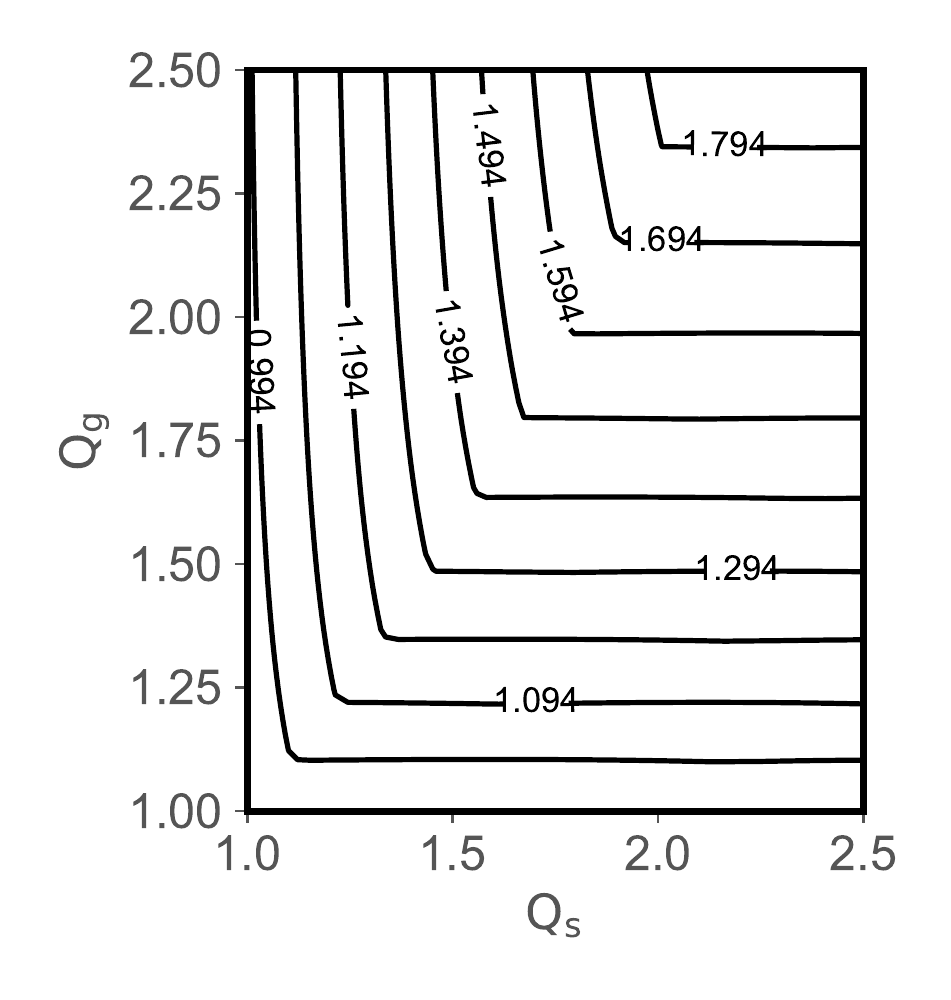}}
\resizebox{65mm}{60mm}{\includegraphics{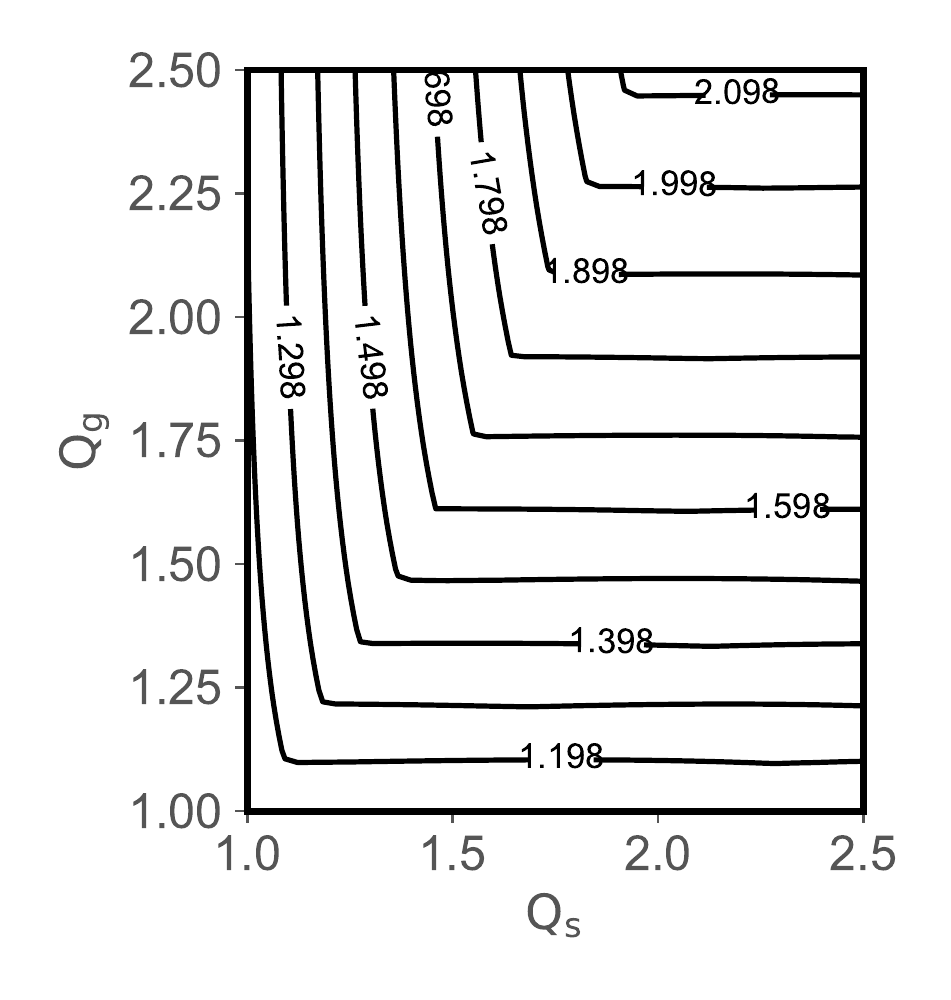}} 
\end{tabular}
\end{center}
\caption{The above plots indicate contours of $Q_{T}$ plotted against $Q_{g}$ and $Q_{s}$, at constant value of $\epsilon =0.05$. 
Panel 1 indicates stability when $R=0$, panel 2 for disruptive tidal field $\rm  R=0.5$}
\end{figure*}

\begin{figure*}
\begin{center}
\begin{tabular}{cc}
\resizebox{65mm}{60mm}{\includegraphics{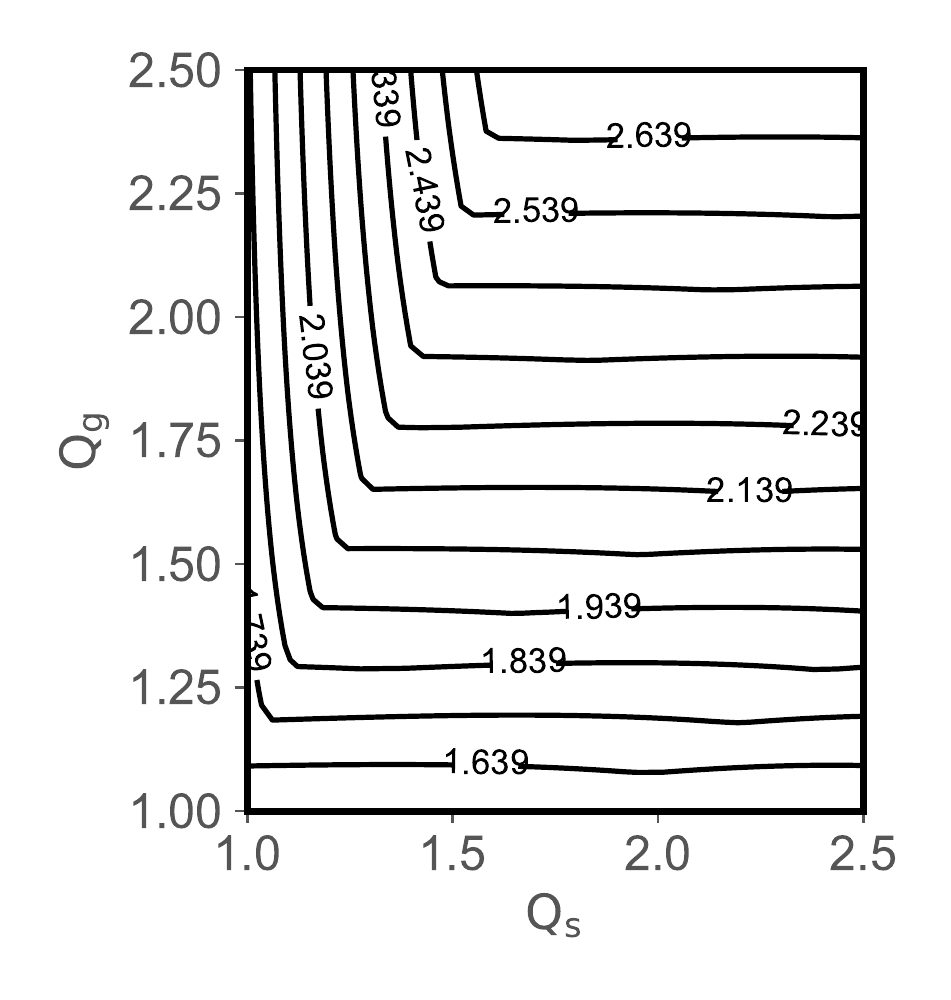}} 
\resizebox{65mm}{60mm}{\includegraphics{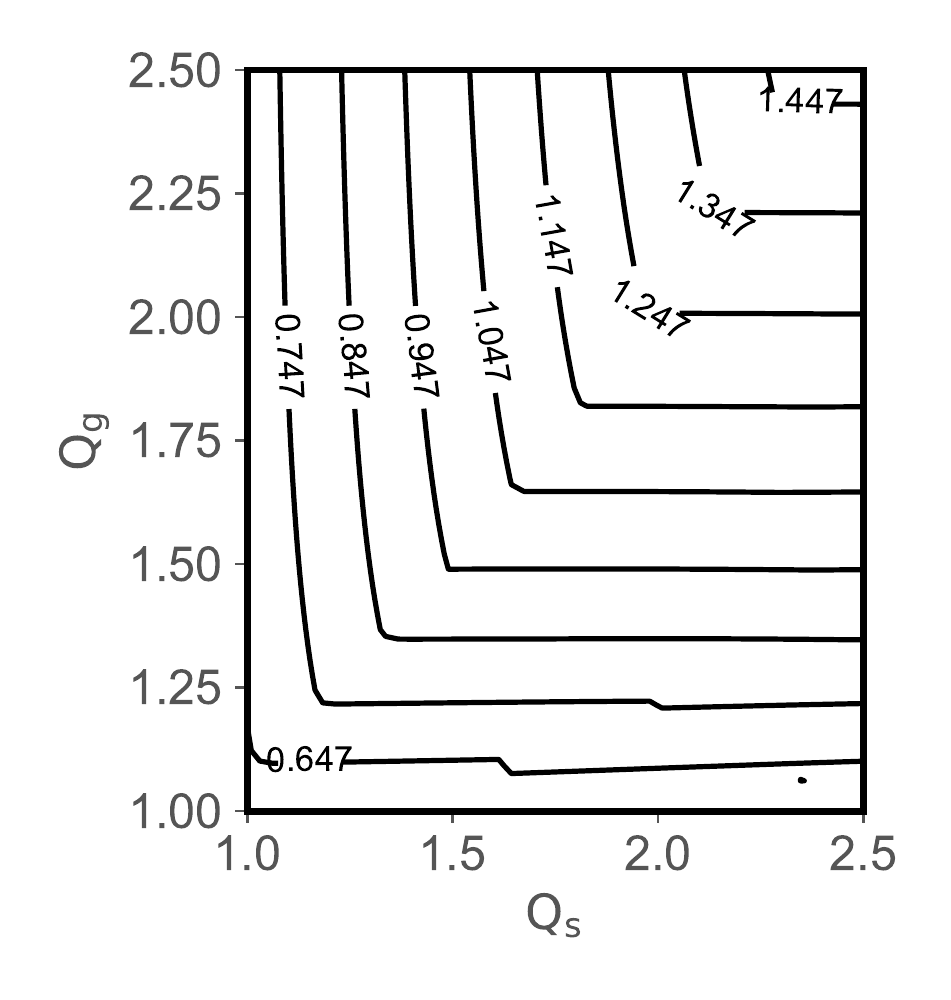}}
\end{tabular}
\end{center}
\caption{The above plots indicate contours of $Q_{T}$ plotted against $Q_{g}$ and $Q_{s}$, at constant value of $\epsilon =0.05$. 
Panel 1 indicates stability at $\rm R=+2.5$ and panel 2 for compressive tidal field $\rm(R=-0.5)$..}
\end{figure*}

\begin{figure*}
\begin{center}
\begin{tabular}{cc}
\resizebox{65mm}{60mm}{\includegraphics{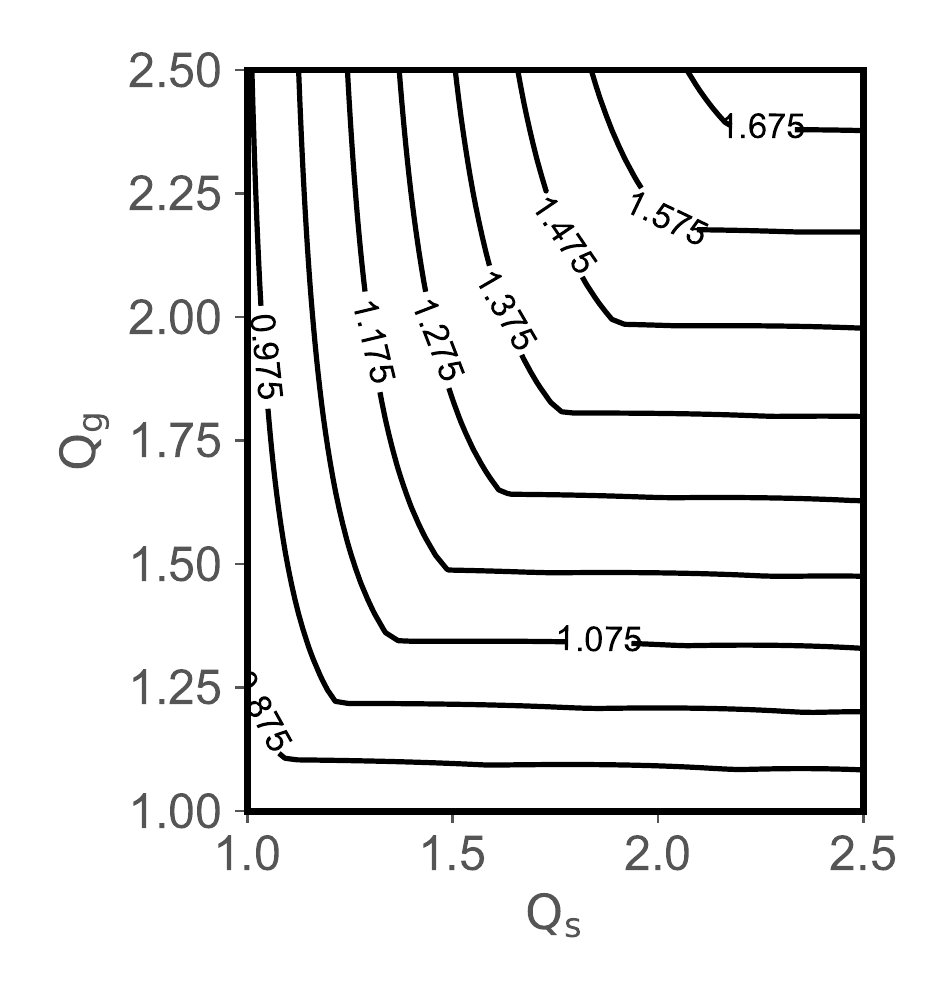}}
\resizebox{65mm}{60mm}{\includegraphics{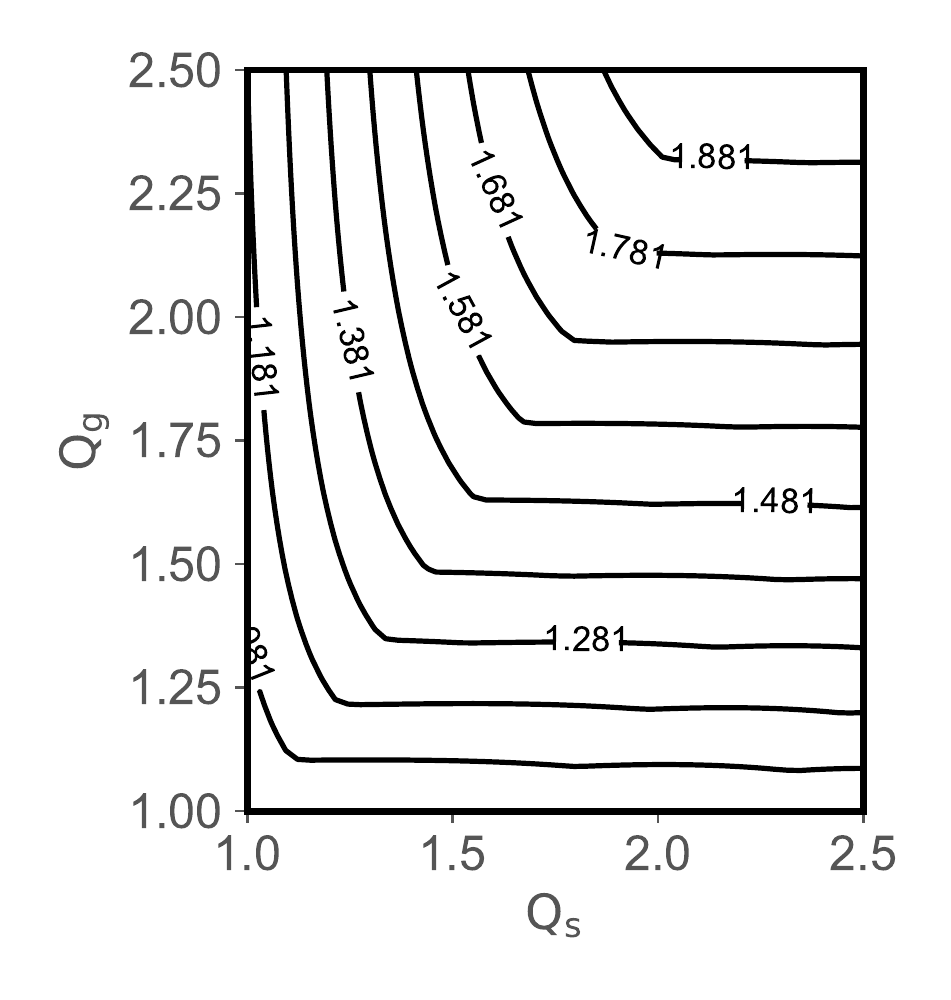}} 
\end{tabular}
\end{center}
\caption{The above plots indicate contours of $Q_{T}$ plotted against $Q_{g}$ and $Q_{s}$, at constant value of $\epsilon =0.1$. 
Panel 1 indicates stability when $R=0$, panel 2 for disruptive tidal field $\rm R=+0.5$}
\end{figure*}

\begin{figure*}
\begin{center}
\begin{tabular}{cc}
\resizebox{65mm}{60mm}{\includegraphics{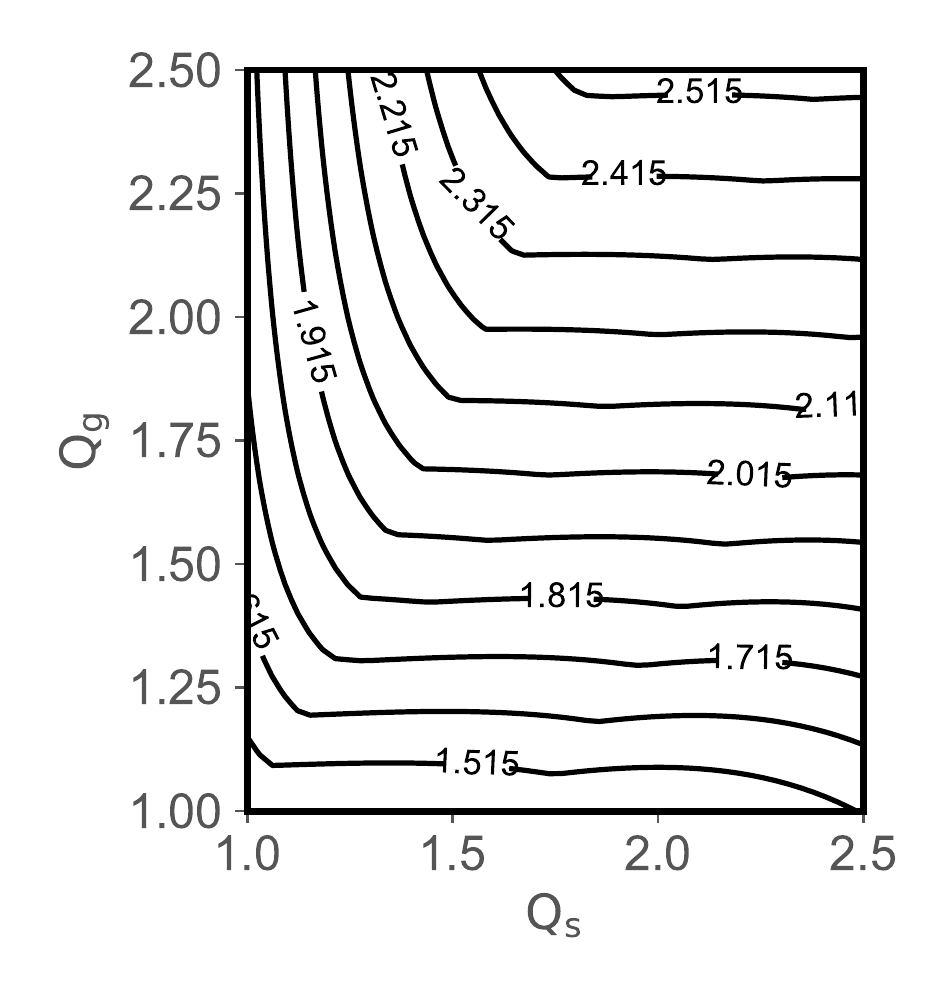}} 
\resizebox{65mm}{60mm}{\includegraphics{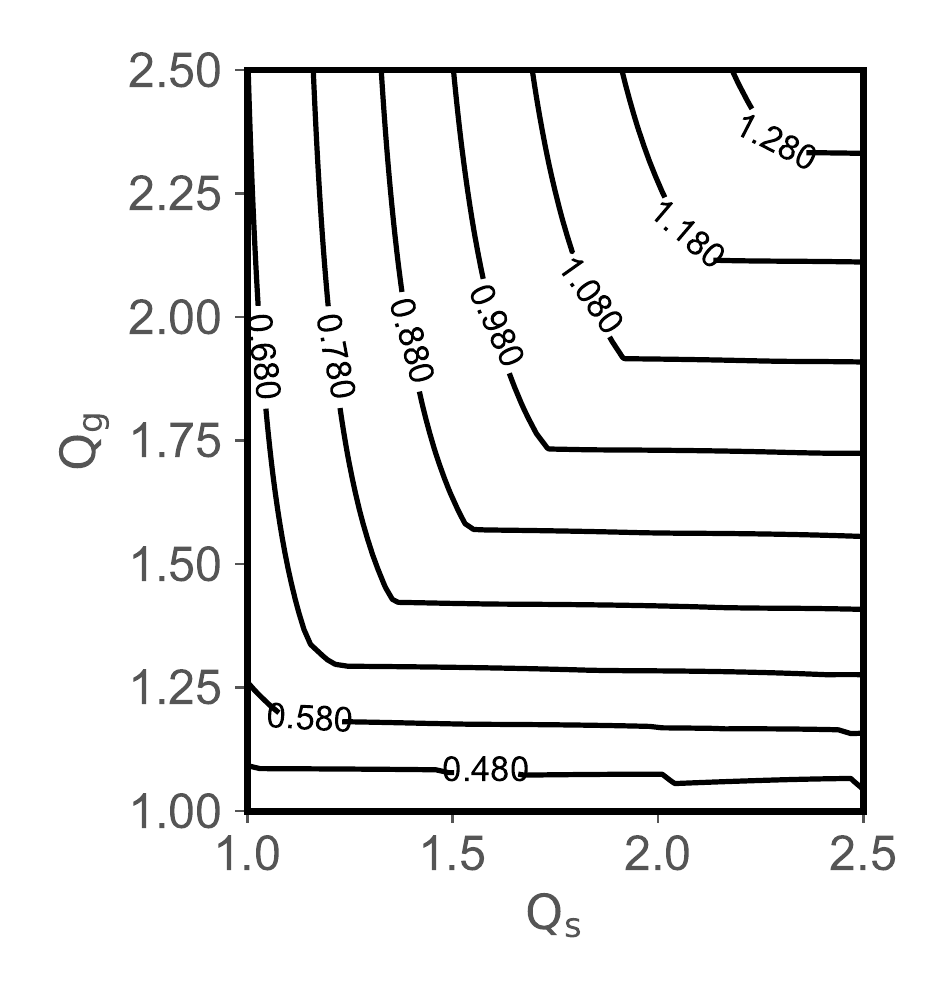}}
\end{tabular}
\end{center}
\caption{The above plots indicate contours of $Q_{T}$ plotted against $Q_{g}$ and $Q_{s}$, at constant value of $\epsilon =0.1$. 
Panel 1 inducates stability at $\rm (R=+2.5)$ and panel 2 for compressive tidal field $\rm R=-0.5$}
\end{figure*}

\begin{figure*}
\begin{center}
\begin{tabular}{cc}
\resizebox{65mm}{60mm}{\includegraphics{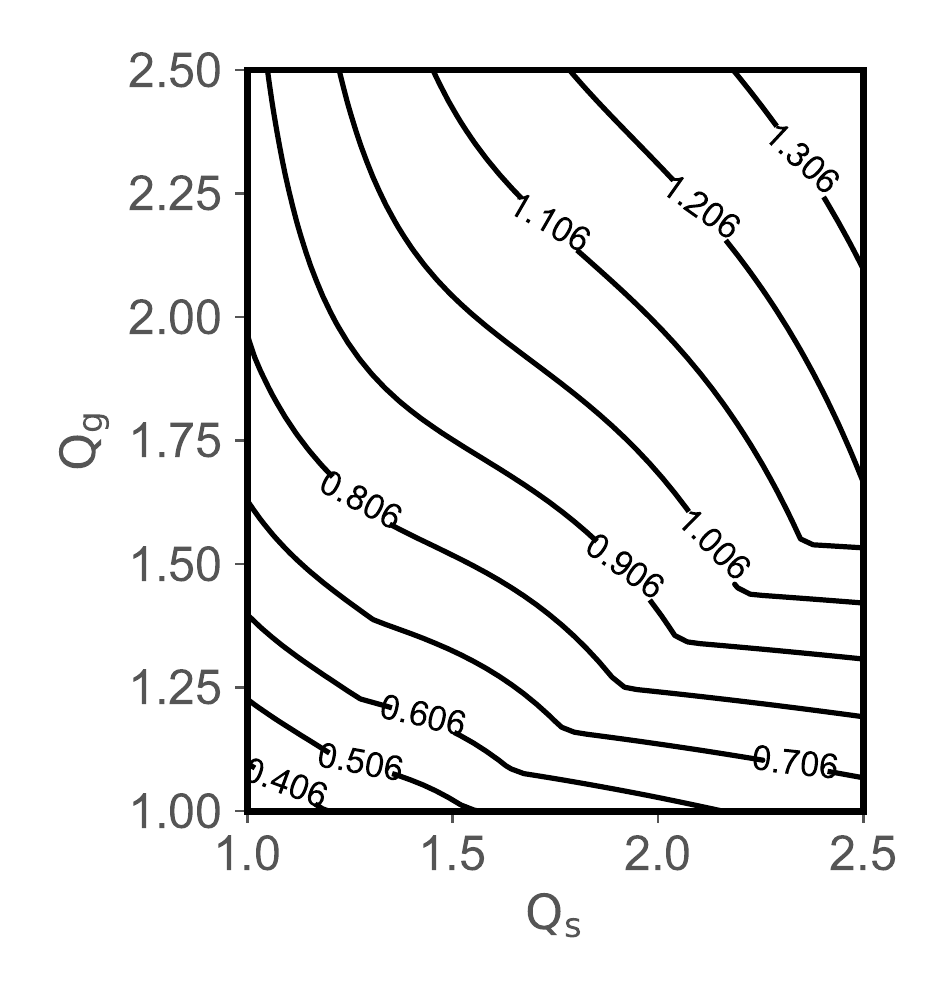}}
\resizebox{65mm}{60mm}{\includegraphics{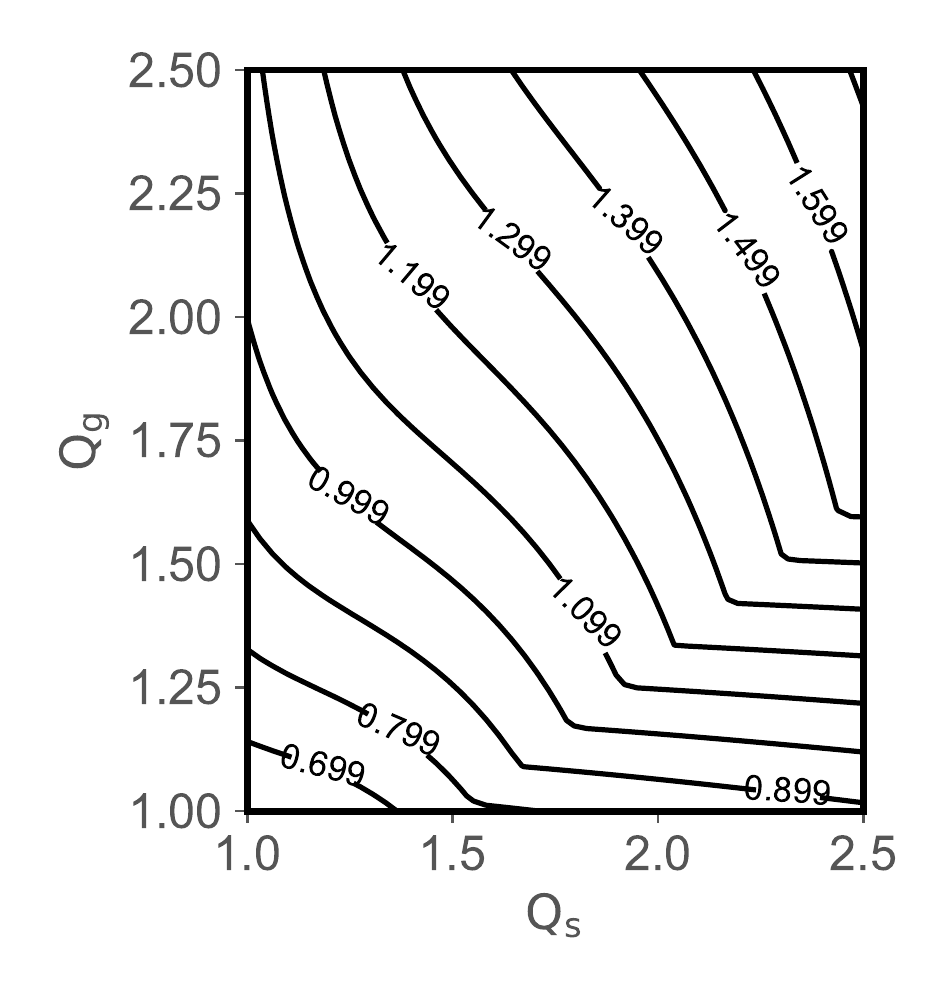}}
\end{tabular}
\end{center}
\caption{The above plots indicate contours of $Q_{T}$ plotted against $Q_{g}$ and $Q_{s}$, at constant value of $\epsilon =0.3$. 
Panel 1 indicates stability when $\rm R=0$, panel 2 for disruptive tidal field $\rm R= +0.5)$.}
\end{figure*}

\begin{figure*}
\begin{center}
\begin{tabular}{ccc}
\resizebox{65mm}{60mm}{\includegraphics{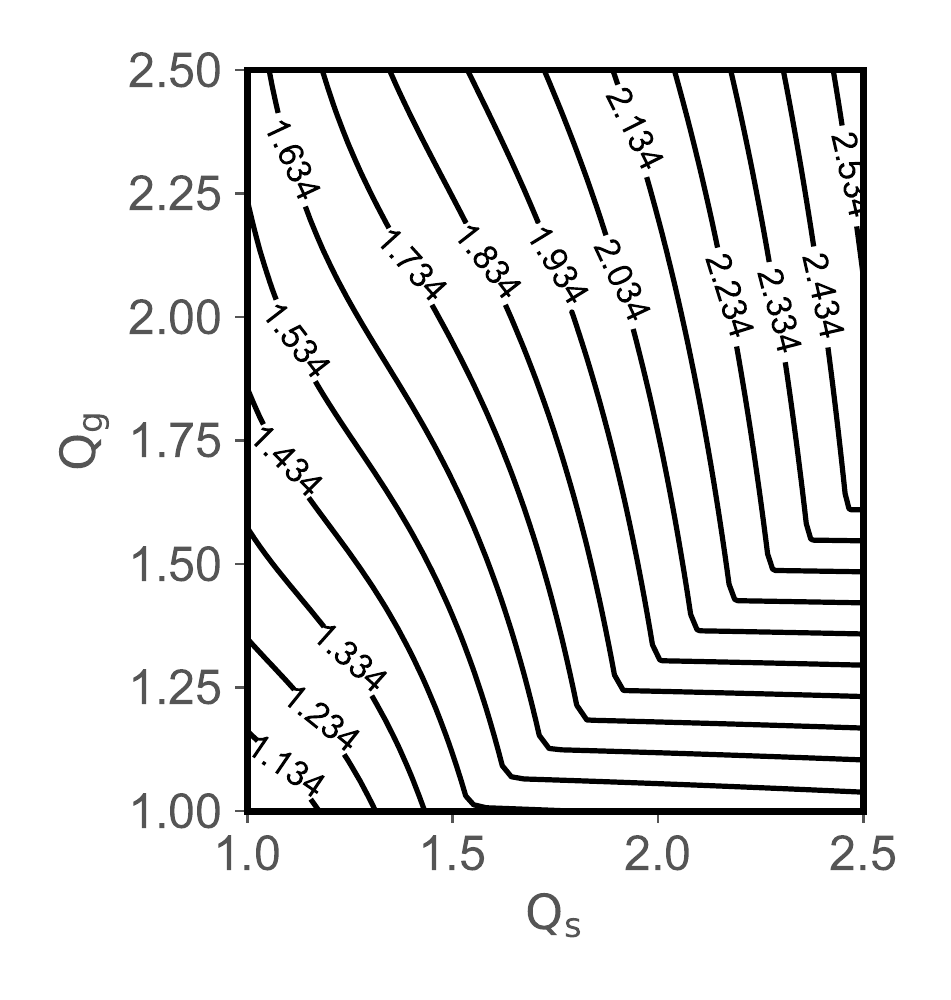}} 
\resizebox{65mm}{60mm}{\includegraphics{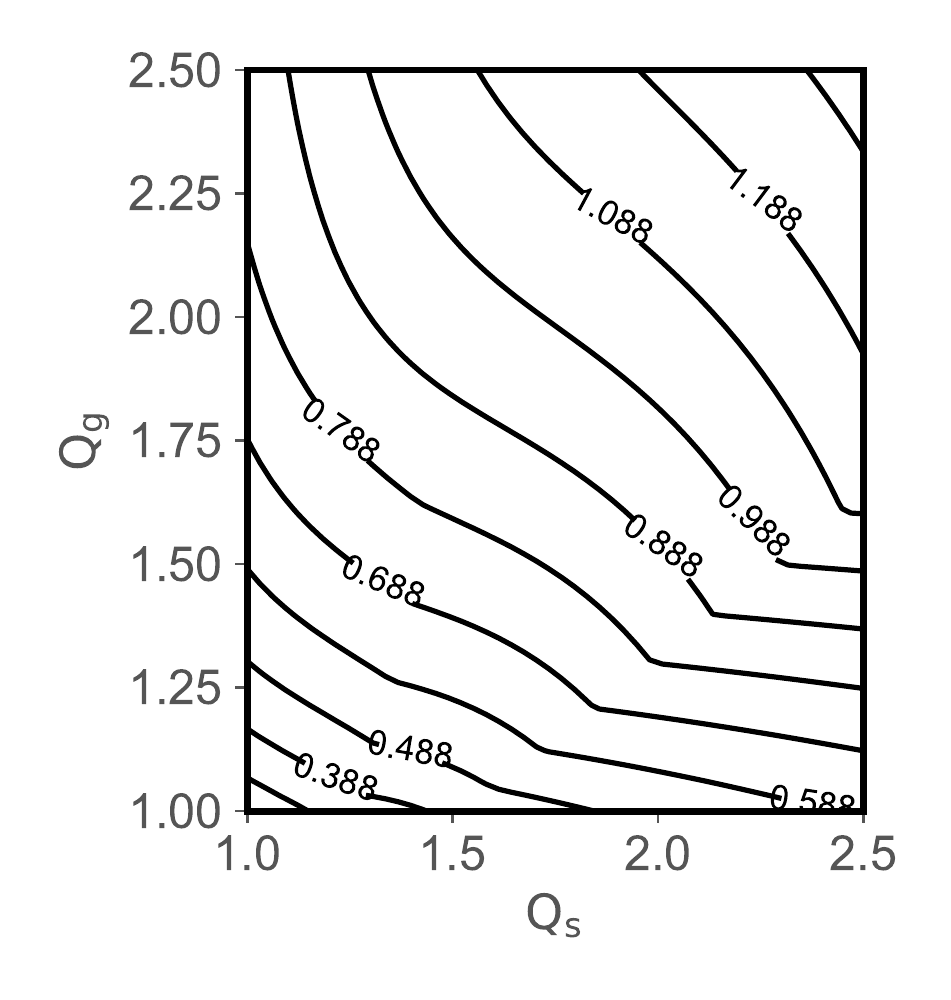}}
\end{tabular}
\end{center}
\caption{The above plots indicate contours of $Q_{T}$ plotted against $Q_{g}$ and $Q_{s}$, at constant value of $\epsilon =0.3$. 
 panel 1 indicates stability for $\rm R=+2.5$ and  panel 2 for compressive tidal field $\rm R=-0.1)$.}
\end{figure*}

The main results of the work can be inferred directly from the table; \\
$\rm 1)$ Increasing gas fraction destabilises the disc, \\$\rm 2)$ A disruptive tidal field tends to 
increase the stability of the disc even at high gas-fraction, whereas compressive tidal field destabilises the galaxy disc.\\

It can be see that by increasing the value of $\epsilon$ from 0.05 to 0.3 changes $Q_{Tmax}$ from 1.794 to 1.306. By applying a disruptive tidal field of 
intermediate strength $R=+0.5$ the value of $Q_{Tmax}$ for $T_{0}=0$ increases from 1.79 to 2.098 and for an intense tidal field $R=2.5$ the  maximum value of stability 
$Q_{T}$ increases to 2.69, which is much higher than either of just stellar or gas disc. Even at very high gas fraction when the disc by itself is unstable 
the disruptive  tidal forces tends to stabilise the disc as an example (see table 2) at a gas fraction of 0.3 without tidal forces the values of lowest stability contour 
is 0.406 and application of $R =+2.5$  makes the disk just stable $Q_{Tmin}=1.012$. Compressive tidal force has a destabilising effect on the galactic disc, it can be seen that
even a tidal field of intermediate strength $R=-0.1$ can destabilise the disk at high gas fraction, see table 1, at gas fraction of about 0.3 when $R=0$,  $Q_{Tmax}=1.306$ 
and  at $R=-0.1$,  $Q_{Tmax}=1.188$. And similarly when $T=0$, for gas fraction $\epsilon=0.3$, $Q_{Tmin}=0.406$ and under influence of an external tidal field 
$R=-0.1$, $Q_{Tmin}=0.388$. 

As a simple possible application we will study the impact of gas-fraction and tidal forces on regulating the growth rate of the instabilities ($\omega_{-}^{2}$) 
in the two-fluid disc. The growth rate of instabilities in two-fluid galactic disc is (see equation 19);

\begin{equation}
\begin{aligned}
\omega^{2}_{-}=\frac{1}{2}(\alpha_{s} + \alpha_{g}) - \frac{1}{2}( (\alpha_{s}+\alpha_{g})^{2} -4(\alpha_{s} \alpha_{g} -\beta_{s} \beta_{g}))^\frac{1}{2}
\end{aligned}
\end{equation}
where (see equation 18)
\begin{equation}
\begin{aligned}
\alpha_{s}= \kappa^{2}  + c_{s} ^{2} k^{2} -  2 \pi G \Sigma_{0s} k + T_{0}\\
\alpha_{g}= \kappa^{2}  + c_{g} ^{2} k^{2} -  2 \pi G \Sigma_{0} k + T_{0}\\
\beta_{s}=2 \pi G \Sigma_{0s} k\\
\beta_{g}=2 \pi G \Sigma_{0g} k
\end{aligned}
\end{equation}

The above quantities can be expressed as 

\begin{equation}
\begin{aligned}
\frac{\alpha_{s}}{\kappa^{2}}= 1+ \frac{1}{4}Q_{s}^{2}(1-\epsilon)^{2} (\frac{k}{k_{T}})^{2} -(1-\epsilon)\frac{k}{k_{T}} + R\\
\frac{\alpha_{g}}{\kappa^{2}}= 1+ \frac{1}{4}Q_{g}^{2}\epsilon^{2} (\frac{k}{k_{T}})^{2} - \epsilon \frac{k}{k_{T}} + R\\
\frac{\beta_{s}}{\kappa^{2}}=  (1-\epsilon)\frac{k}{k_{T}}\\
\frac{\beta_{g}}{\kappa^{2}}=  \epsilon\frac{k}{k_{T}}
\end{aligned}
\end{equation}

Here, $k_{T}=\frac{\kappa^{2}}{2 \pi G (\Sigma_{0s} + \Sigma_{0g})}$, for each case we will fix the values of $Q_{s}$, $Q_{g}$ and the gas fraction $\epsilon$ and vary the value of $R$ to see the effect of the tidal forces on
the two-fluid disc.
\FloatBarrier
\begin{figure*}
\begin{center}
\begin{tabular}{ccccc}
\centering
\resizebox{30mm}{33mm}{\includegraphics{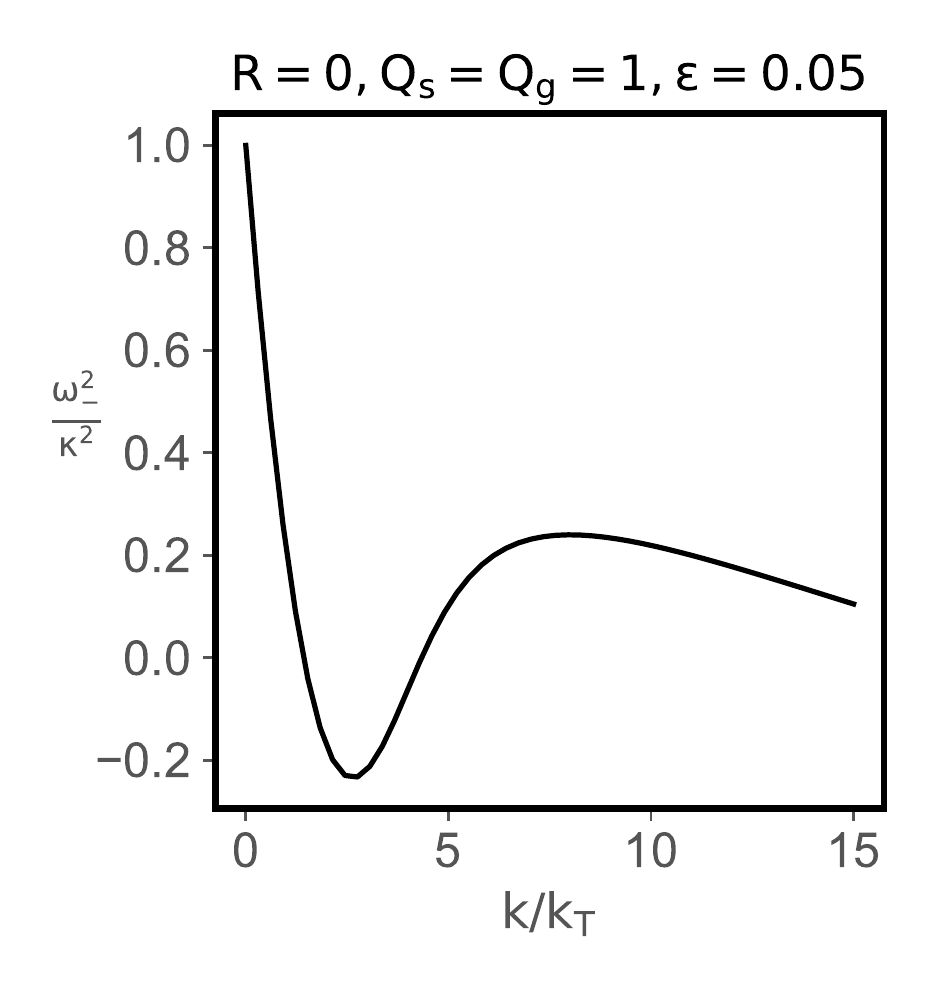}}
\resizebox{30mm}{33mm}{\includegraphics{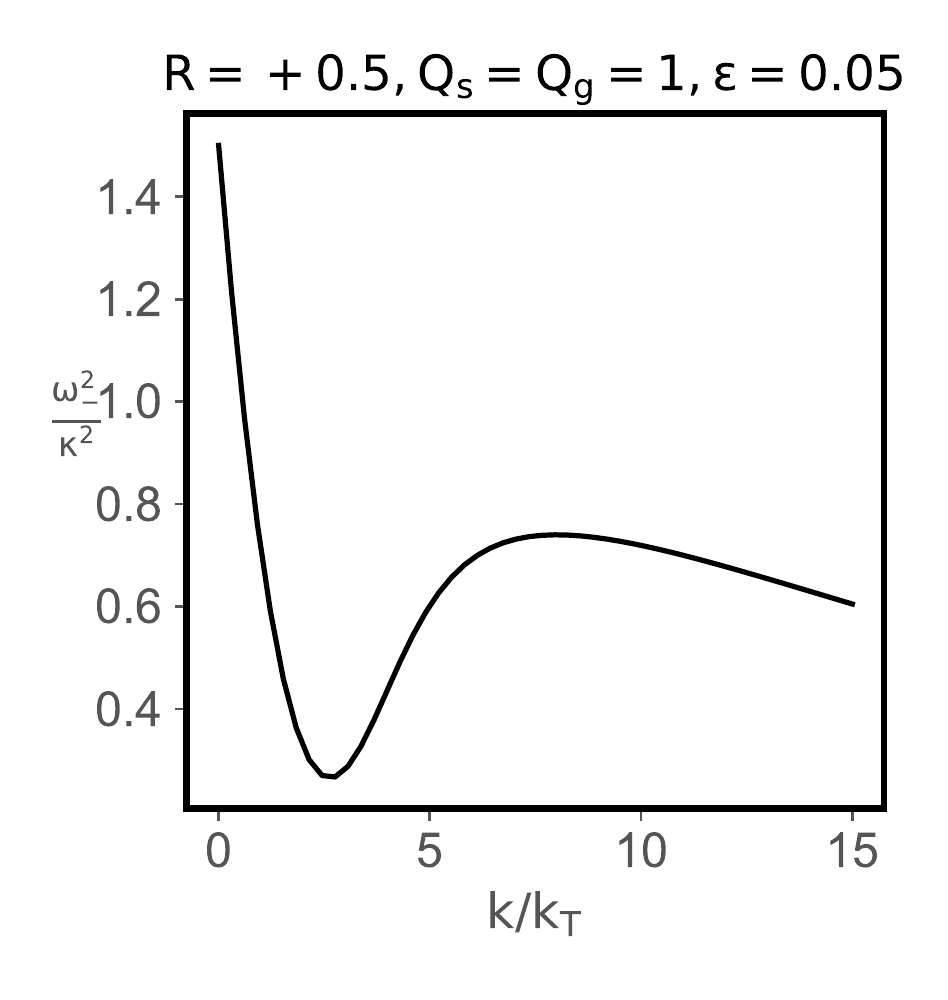}}
\resizebox{30mm}{33mm}{\includegraphics{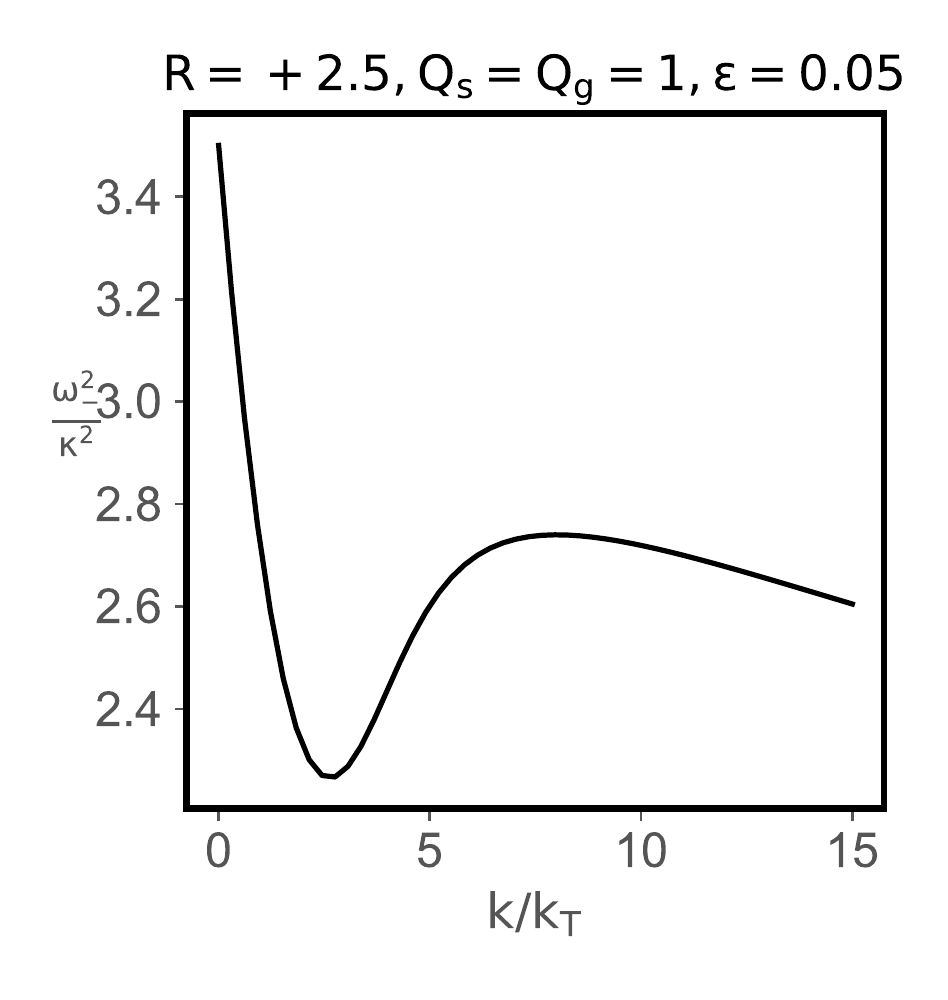}}
\resizebox{30mm}{33mm}{\includegraphics{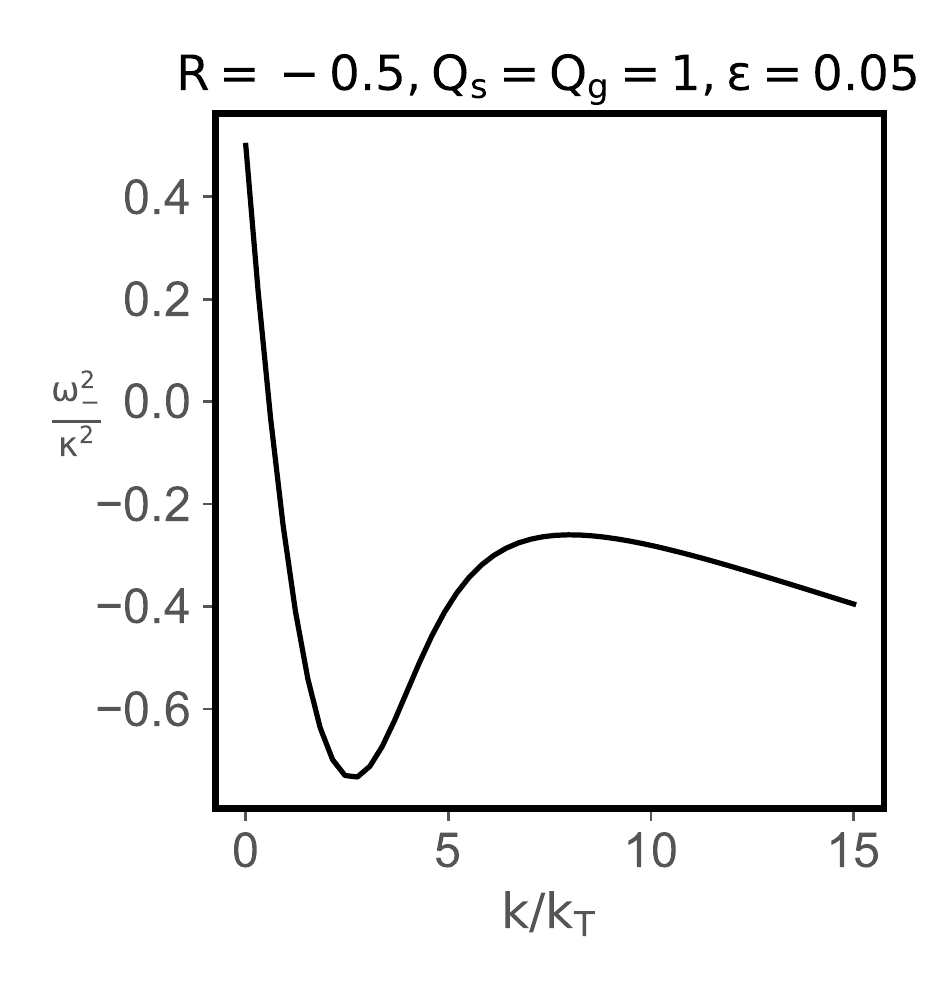}}
\resizebox{30mm}{33mm}{\includegraphics{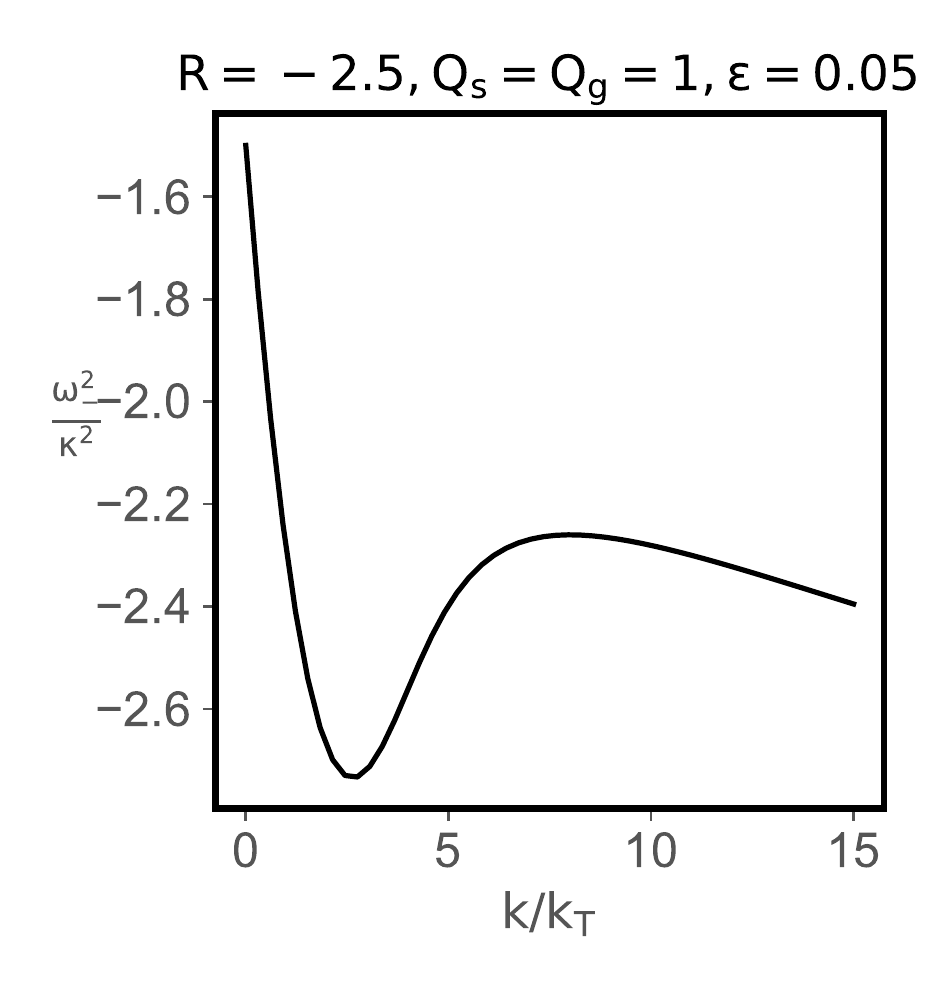}}
\end{tabular}
\end{center}
\caption{The above plots indicate the growth rate $\omega_{-}^{2}$ for varying value of R= 0, +0.5, +2.5, -0.5, -2.5 at a constant value of 
$\epsilon$=0.05 and $Q_{s}$=$Q_{g}$=1. }
\end{figure*}

\begin{figure*}
\begin{center}
\begin{tabular}{ccccc}
\centering
\resizebox{30mm}{33mm}{\includegraphics{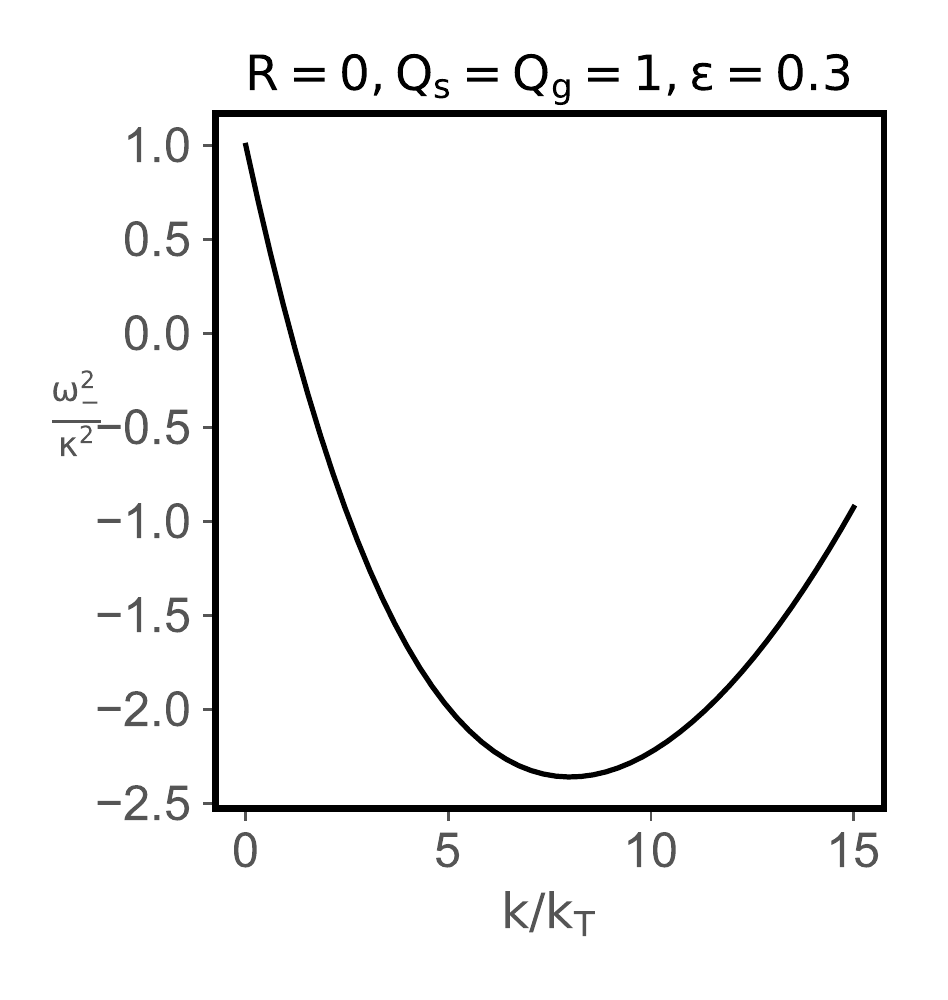}}
\resizebox{30mm}{33mm}{\includegraphics{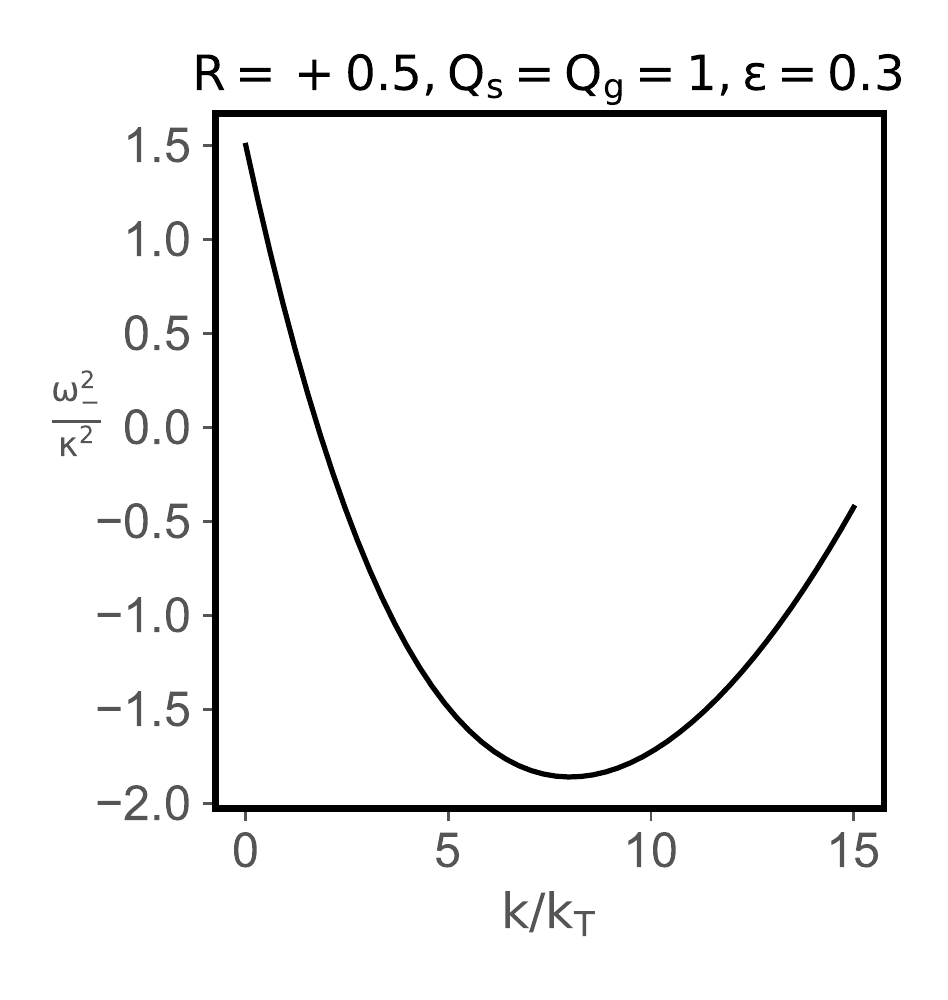}}
\resizebox{30mm}{33mm}{\includegraphics{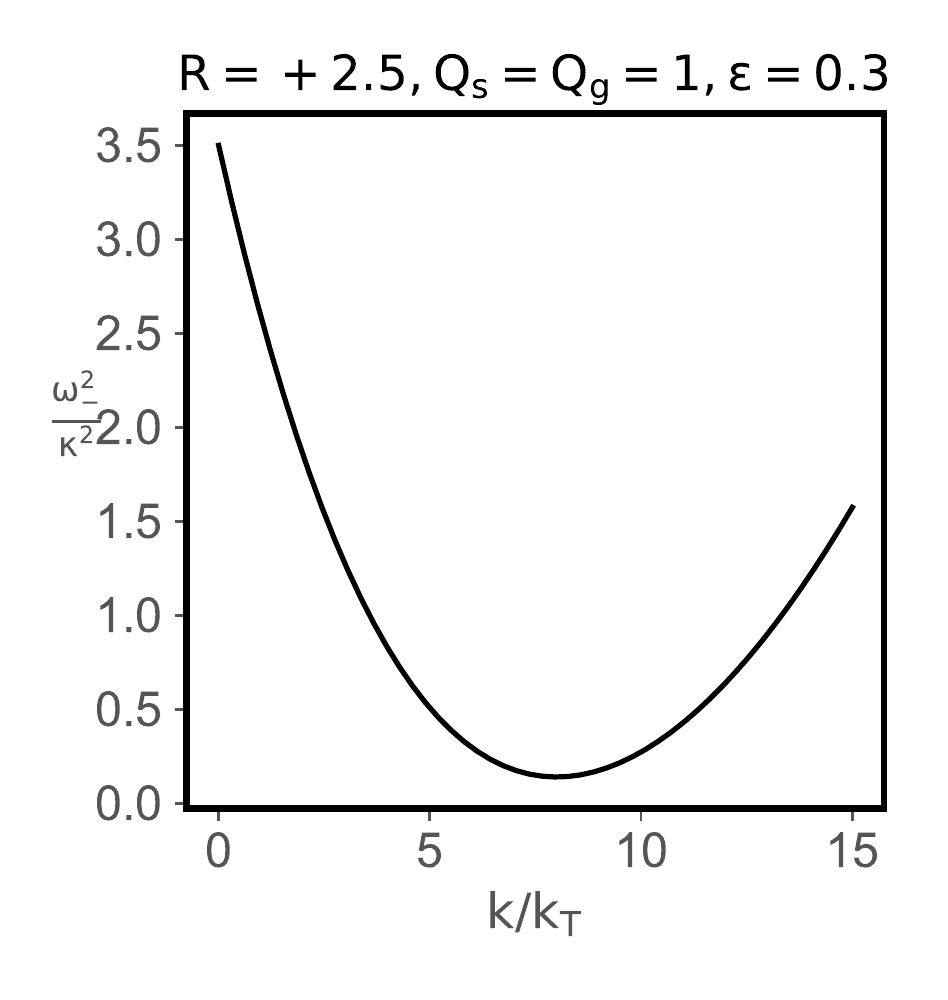}}
\resizebox{30mm}{33mm}{\includegraphics{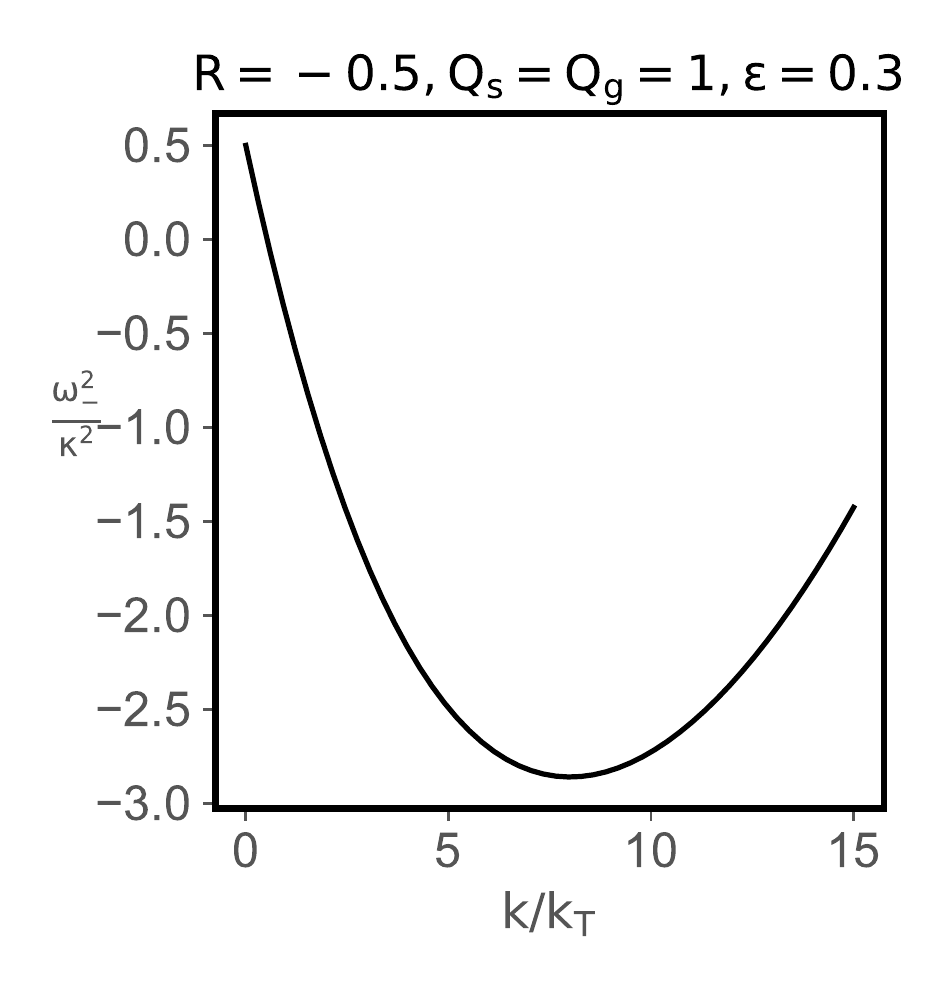}}
\resizebox{30mm}{33mm}{\includegraphics{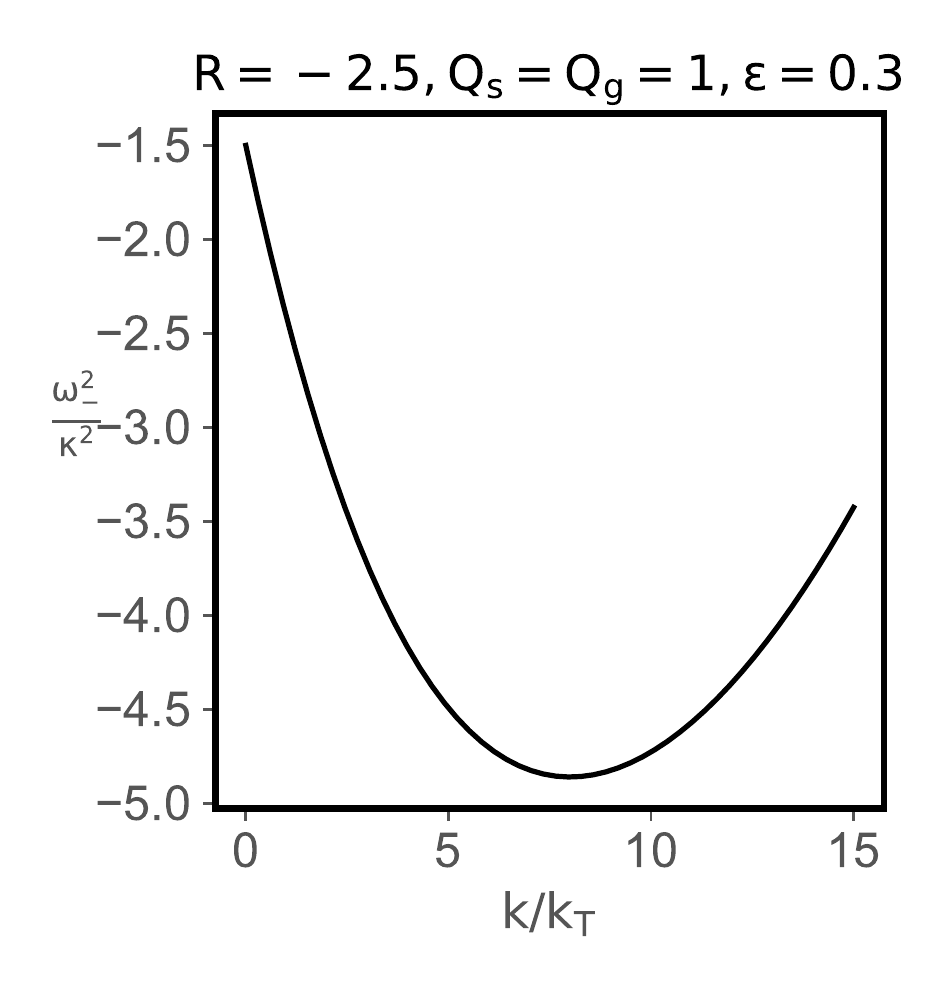}}
\end{tabular}
\end{center}
\caption{The above plots indicate the growth rate $\omega_{-}^{2}$ for varying value of R= 0, +0.5, +2.5, -0.5, -2.5 at a constant value of 
$\epsilon$=0.3 and $Q_{s}$=$Q_{g}$=1. }
\end{figure*}
%========================================================================================================================
\begin{figure*}
\begin{center}
\begin{tabular}{ccccc}
\centering
\resizebox{30mm}{33mm}{\includegraphics{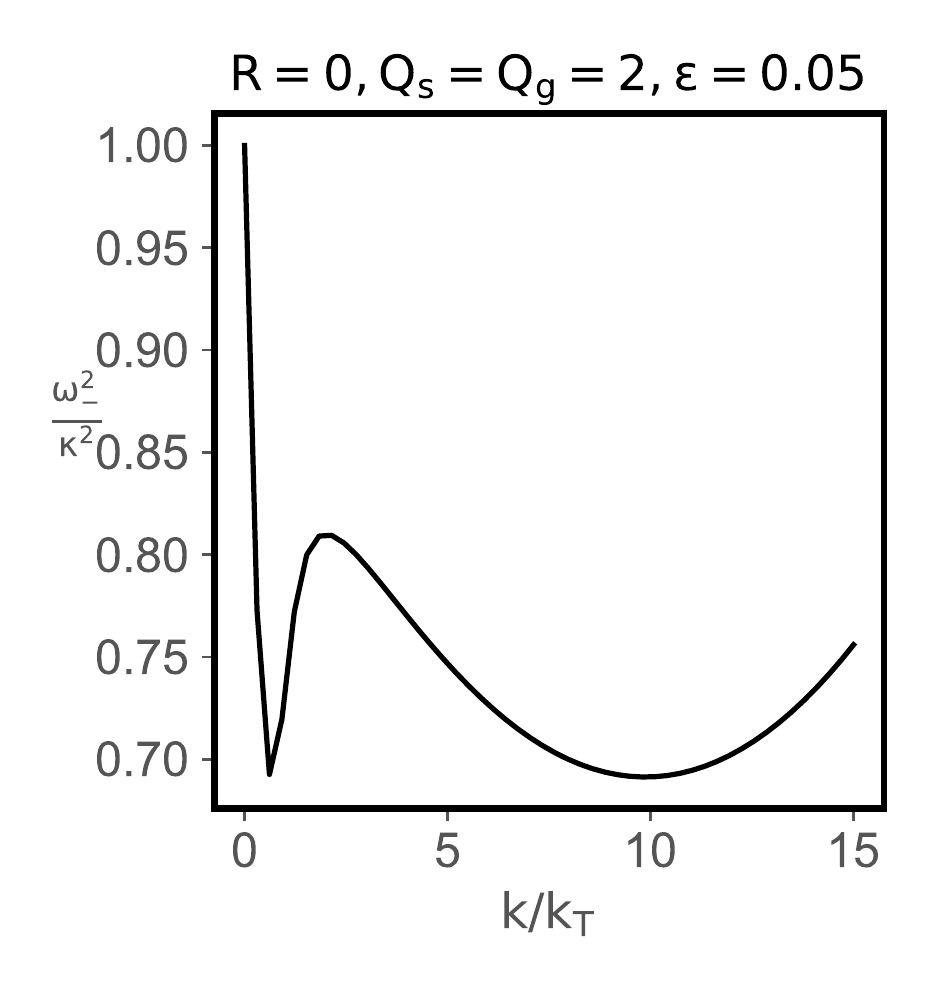}}
\resizebox{30mm}{33mm}{\includegraphics{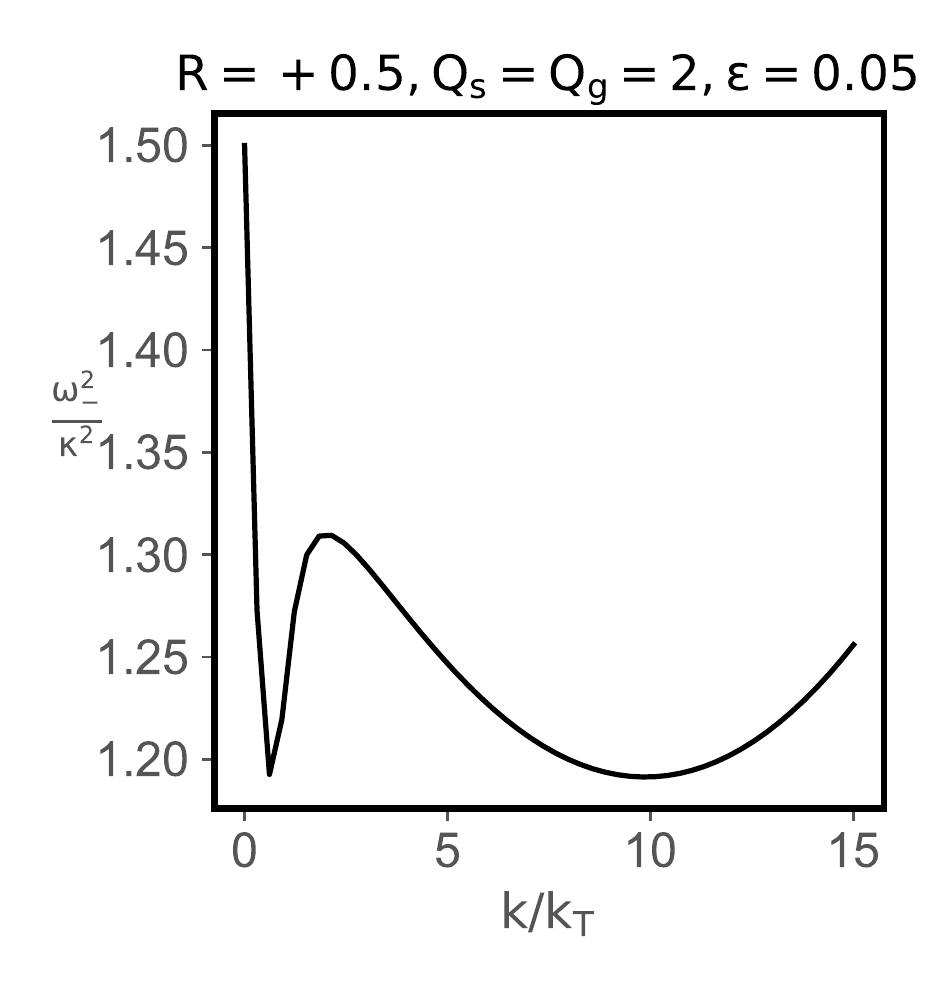}}
\resizebox{30mm}{33mm}{\includegraphics{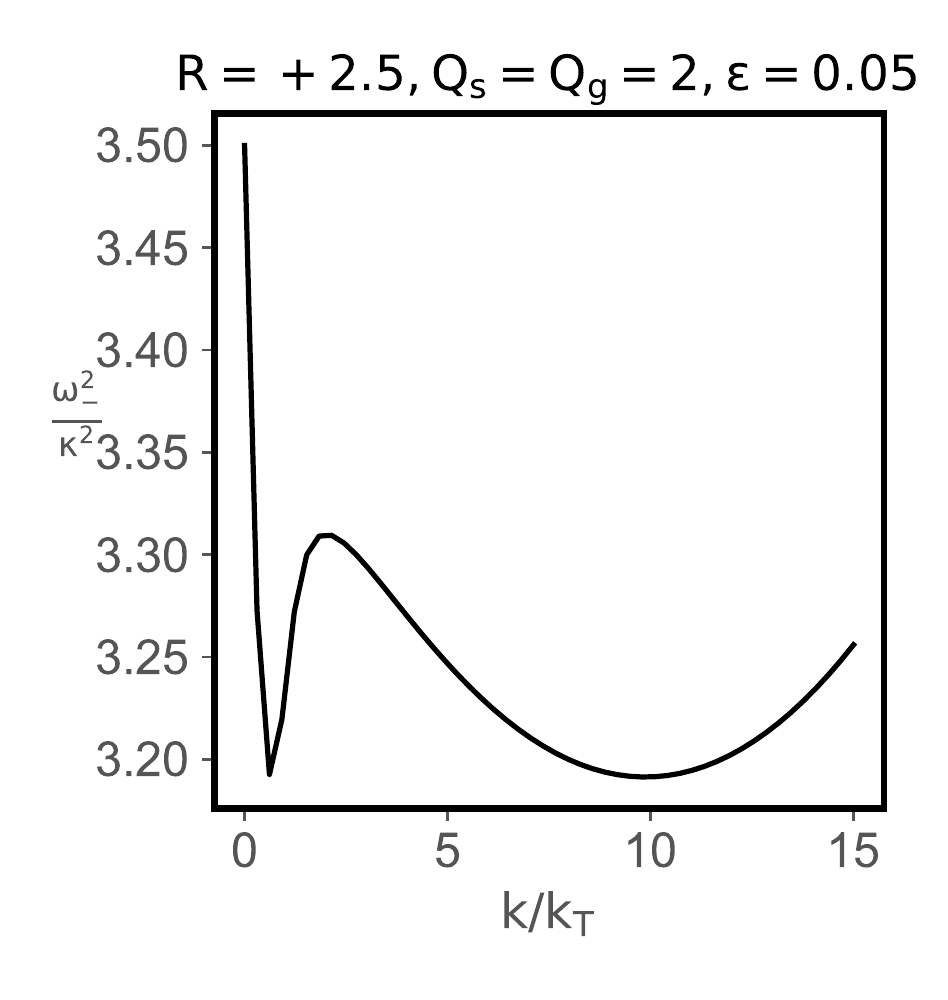}}
\resizebox{30mm}{33mm}{\includegraphics{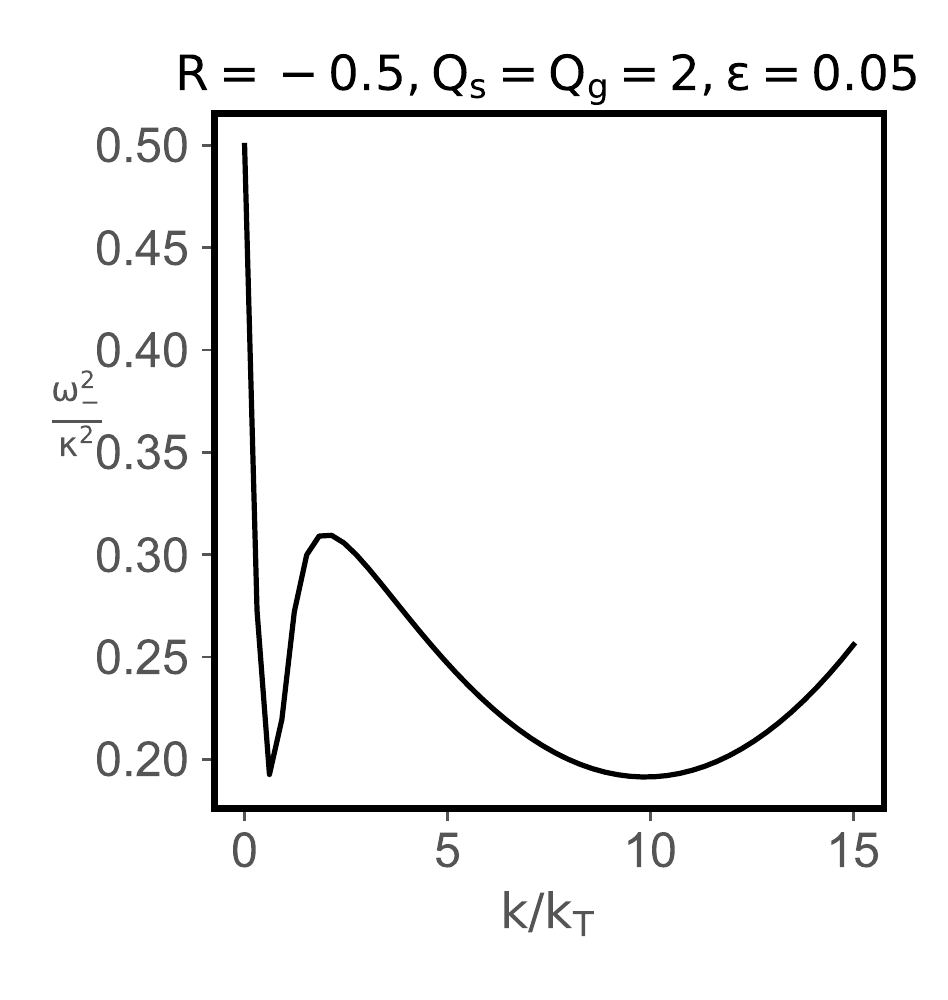}}
\resizebox{30mm}{33mm}{\includegraphics{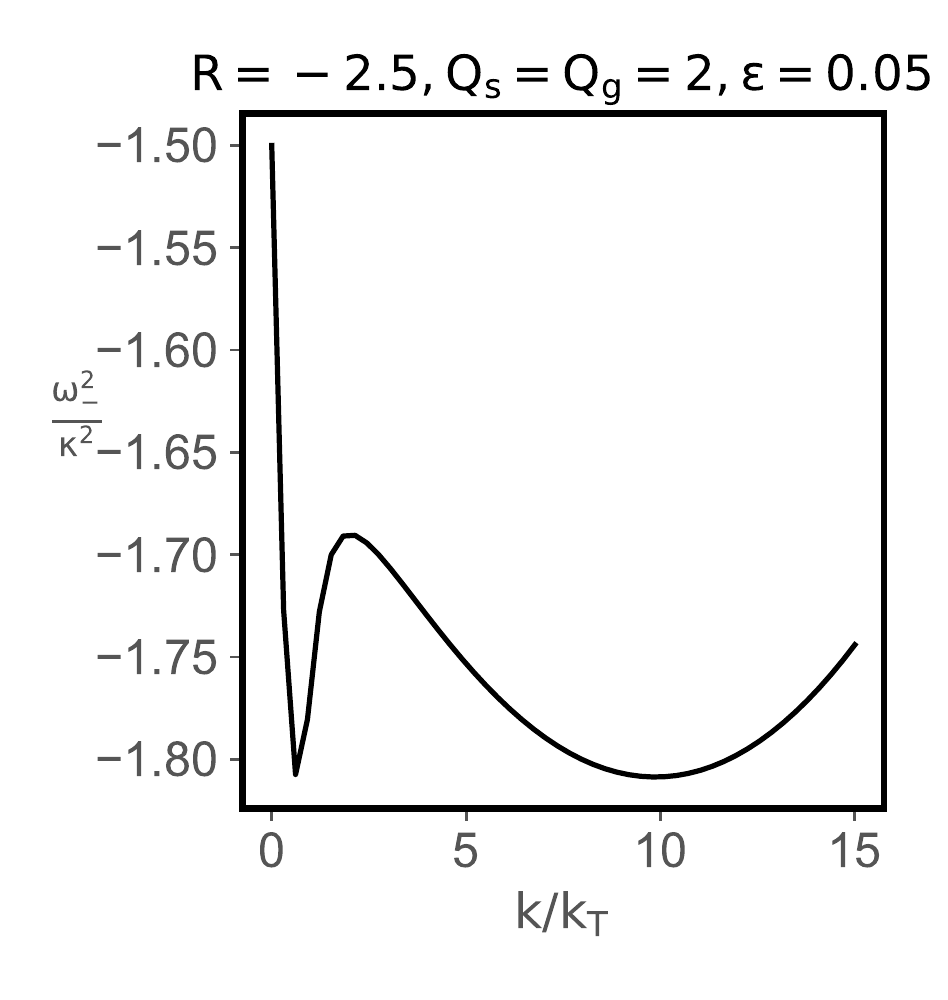}}
\end{tabular}
\end{center}
\caption{The above plots indicate the growth rate $\omega_{-}^{2}$ for varying value of R= 0, +0.5, +2.5, -0.5, -2.5 at a constant value of 
$\epsilon$=0.05 and $Q_{s}$=$Q_{g}$=2. }
\end{figure*}
%===================================================================================================================
\begin{figure*}
\begin{center}
\begin{tabular}{ccccc}
\centering
\resizebox{30mm}{33mm}{\includegraphics{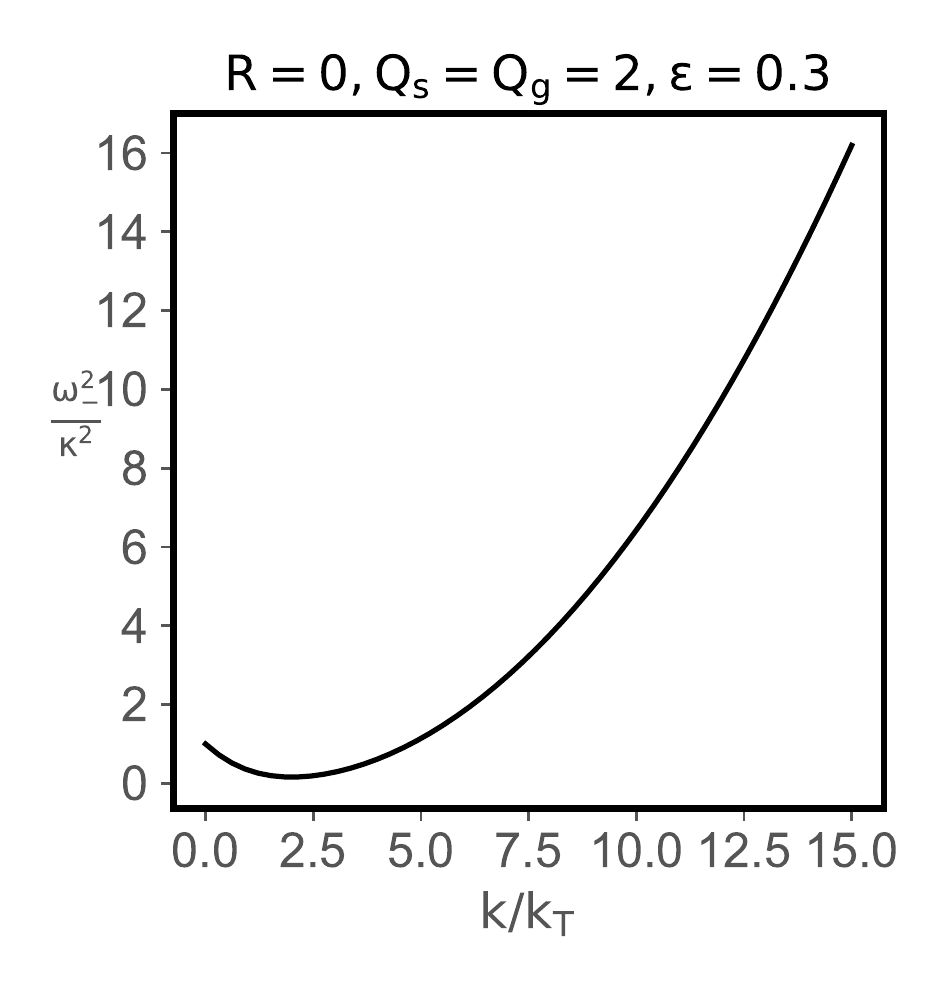}}
\resizebox{30mm}{33mm}{\includegraphics{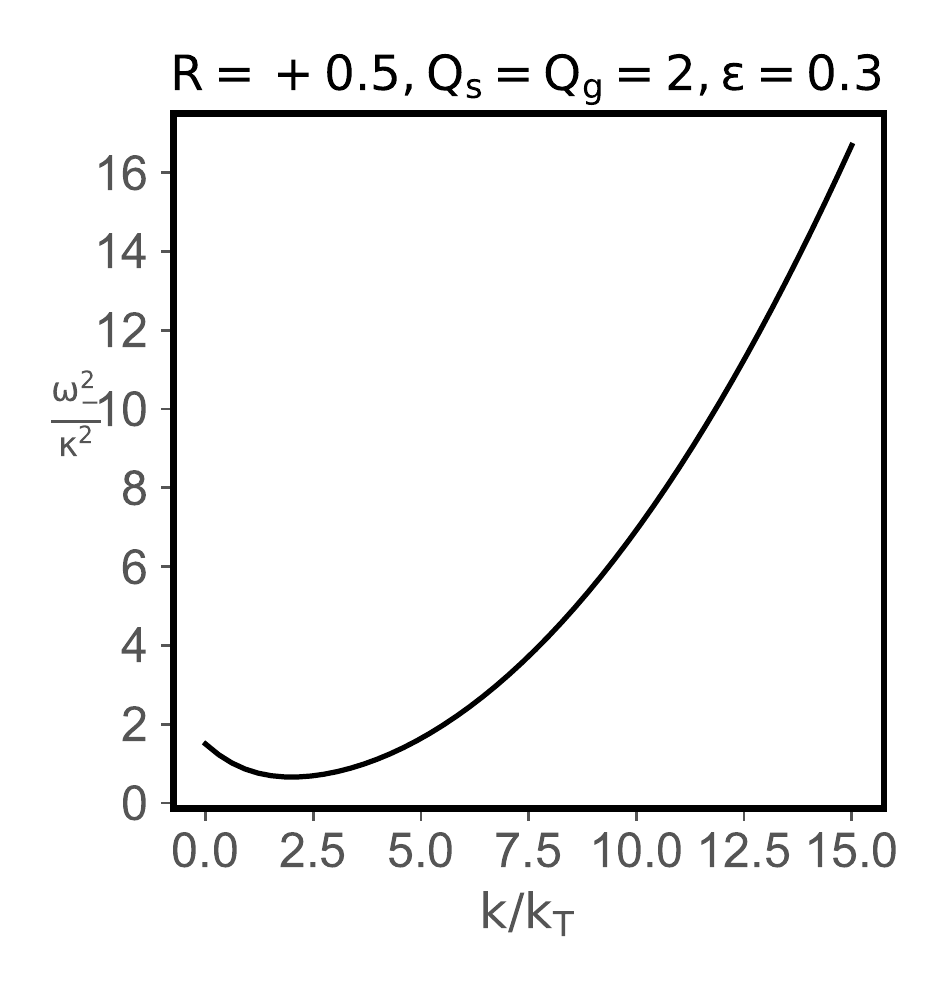}}
\resizebox{30mm}{33mm}{\includegraphics{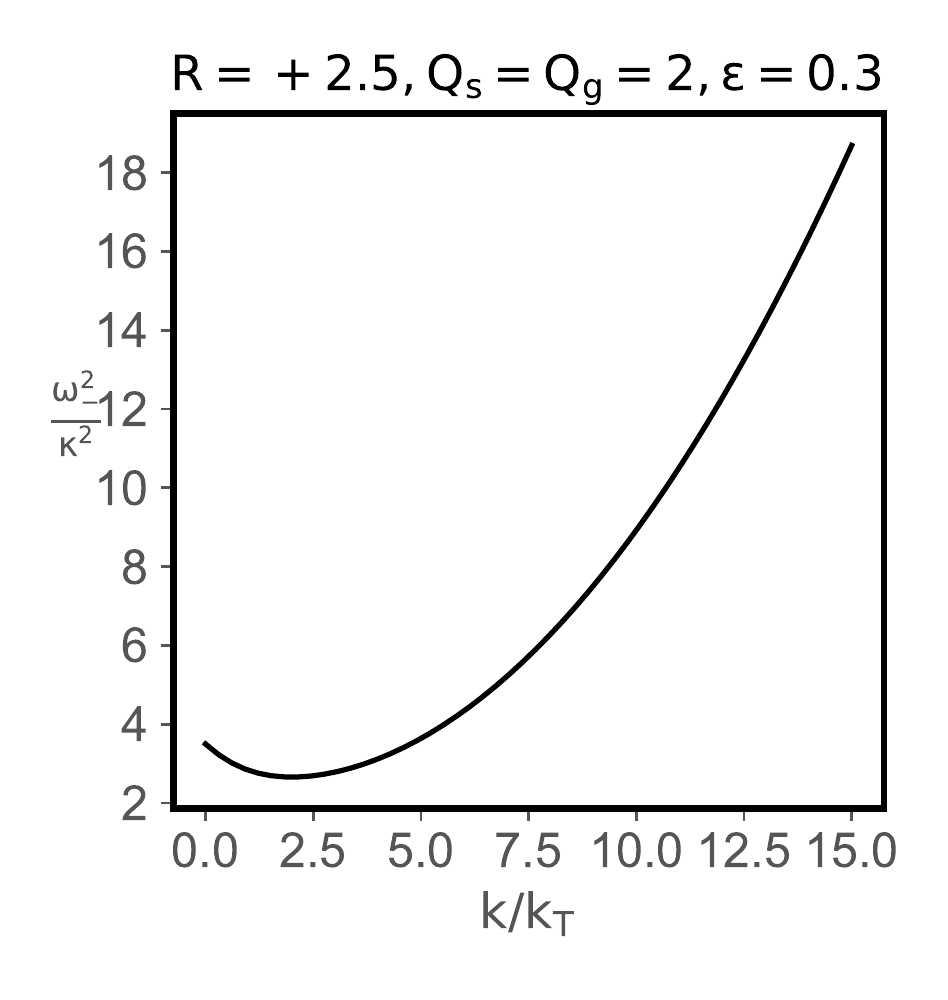}}
\resizebox{30mm}{33mm}{\includegraphics{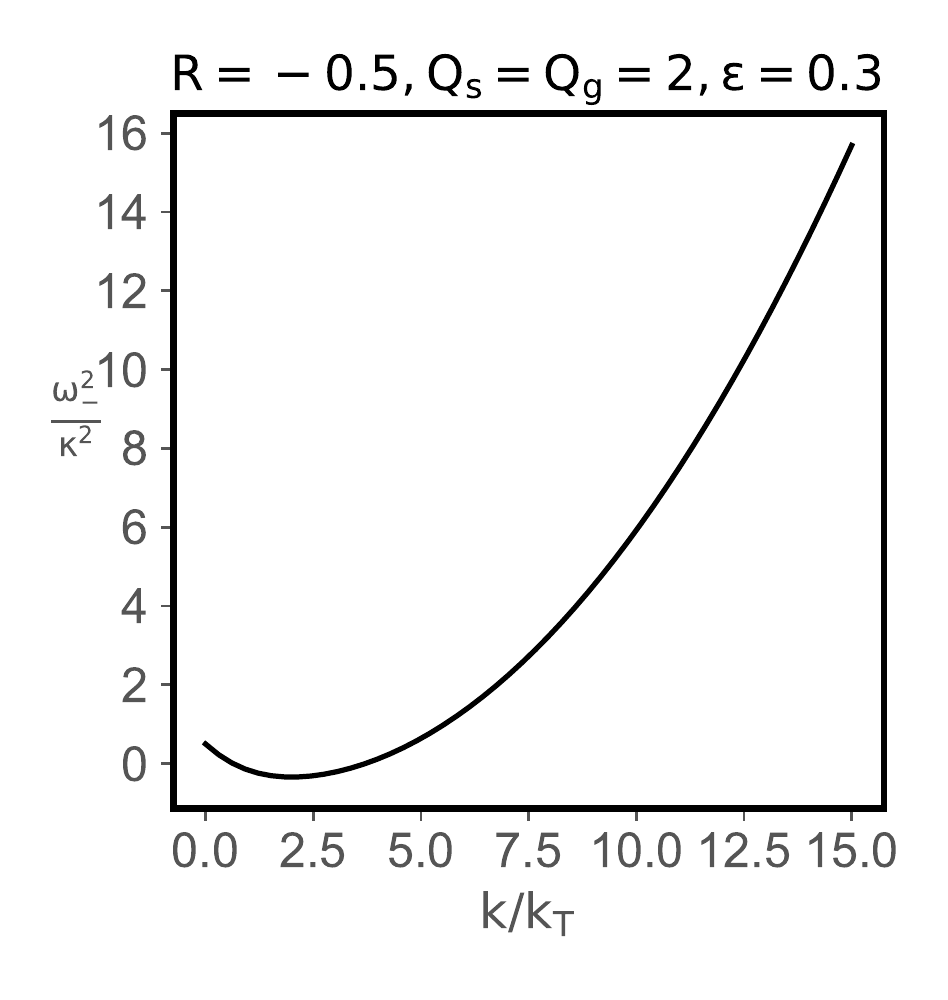}}
\resizebox{30mm}{33mm}{\includegraphics{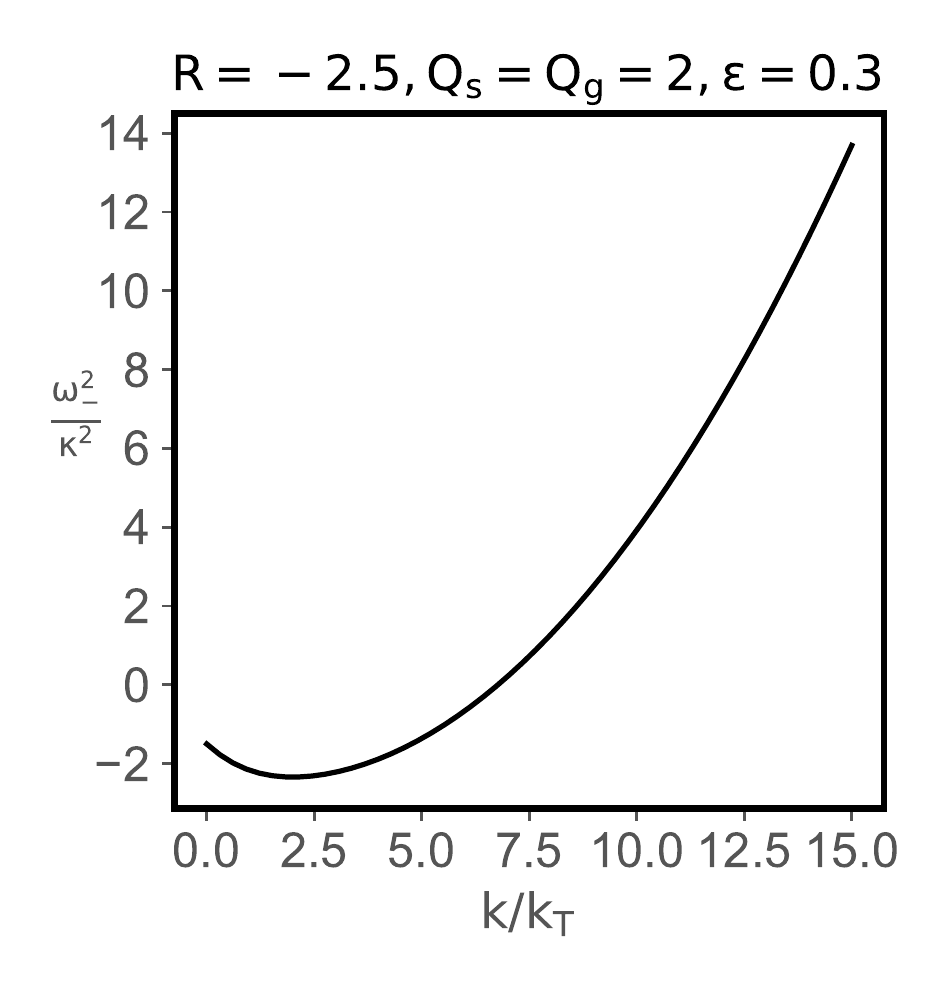}}
\end{tabular}
\end{center}
\caption{The above plots indicate the growth rate $\omega_{-}^{2}$ for varying value of R= 0, +0.5, +2.5, -0.5, -2.5 at a constant value of 
$\epsilon$=0.3 and $Q_{s}$=$Q_{g}$=2. }
\end{figure*}

\begin{table}
\small
\centering
\begin{tabular}{|l|c|c|c|c|c|}
\hline
$R\rightarrow$          & 0.0 & +0.5  & 2.5 & -0.5 & -2.5  \\
$\epsilon \downarrow$   &     &       &     &      &       \\
\hline
\hline
0.05&-0.232  &0.267    &2.26    &-0.73   &-2.73   \\
0.3 & -2.35  &-1.86   &0.140   &-2.85   &-4.85  \\
\hline
\end{tabular}
\caption{Above table depicts the values of $\omega_{-(min)}^{2}$ at constant value of $Q_{s}=Q_{g}=1$.}
\label{table:D}
\end{table}

\begin{table}
\small
\centering
\begin{tabular}{|l|c|c|c|c|c|}
\hline
$R\rightarrow$          & 0.0 & +0.5  & 2.5 & -0.5 & -2.5  \\
$\epsilon \downarrow$   &     &       &     &      &       \\
\hline
\hline
0.05&0.69&1.19   &3.19   &0.19   &-1.80   \\
0.3 &0.16   &0.66   &2.66   &-0.377   &-2.33  \\
\hline
\end{tabular}
\caption{Above table depicts the values of $\omega_{-(min)}^{2}$ at constant value of $Q_{s}=Q_{g}=2$.}
\label{table:C}
\end{table}

\begin{figure*}
\begin{center}
\begin{tabular}{ccccc}
\centering
\resizebox{30mm}{33mm}{\includegraphics{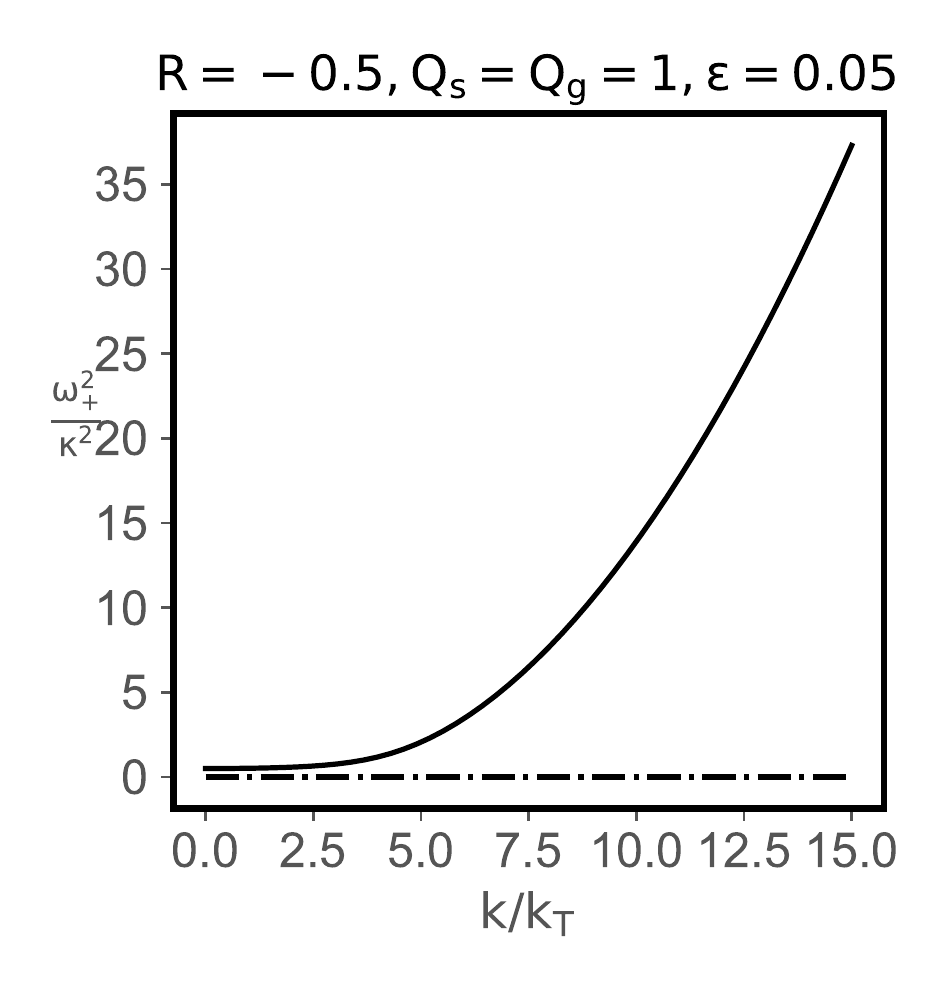}}
\resizebox{30mm}{33mm}{\includegraphics{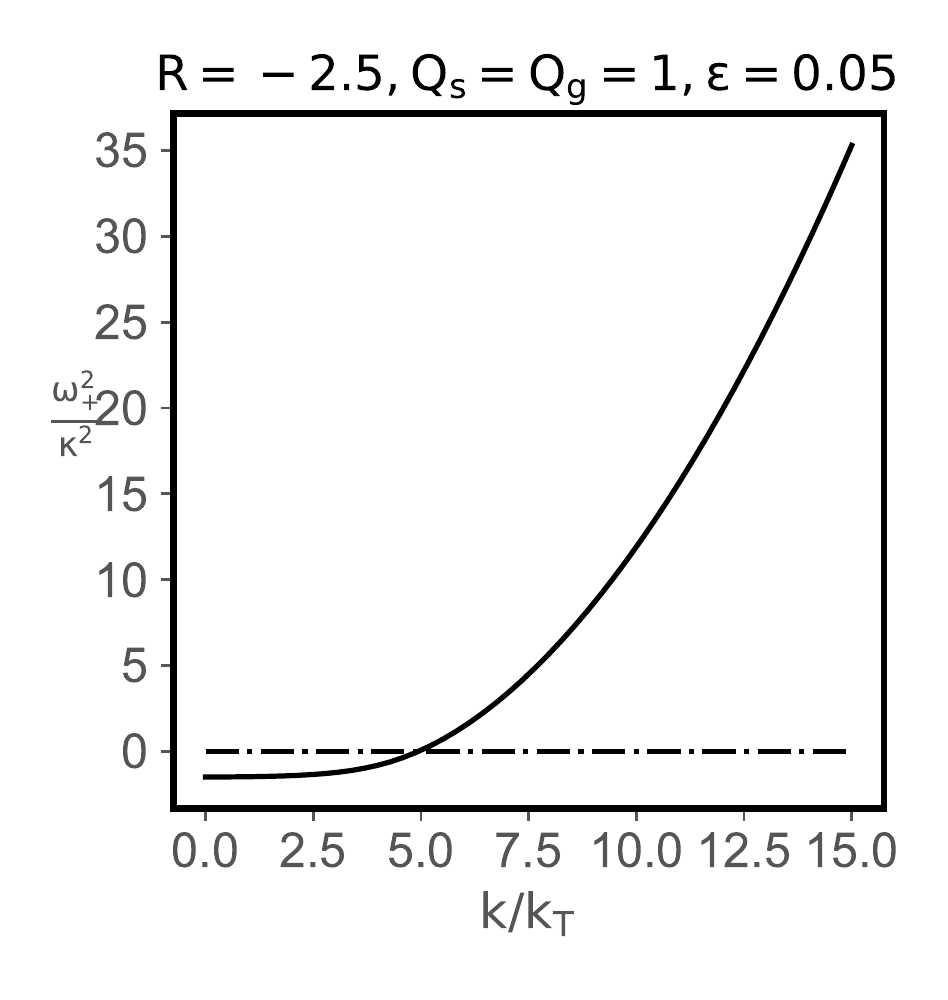}}
\resizebox{30mm}{33mm}{\includegraphics{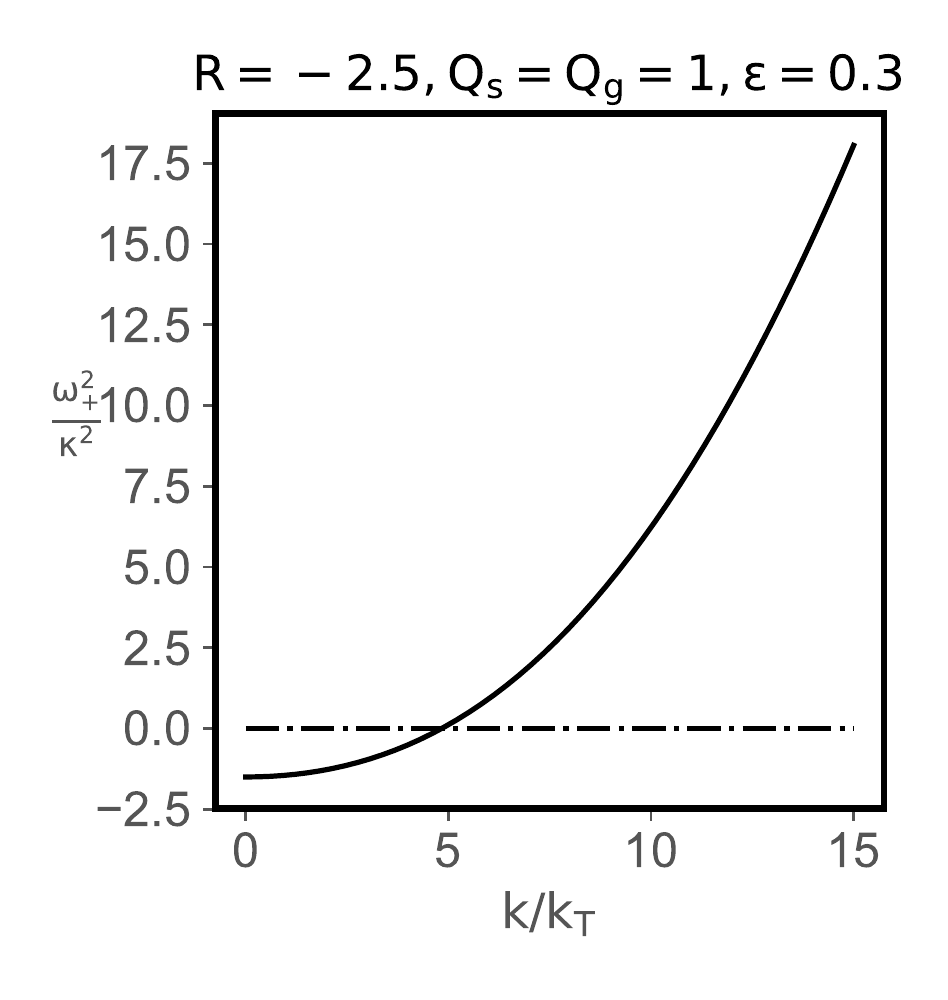}}
\resizebox{30mm}{33mm}{\includegraphics{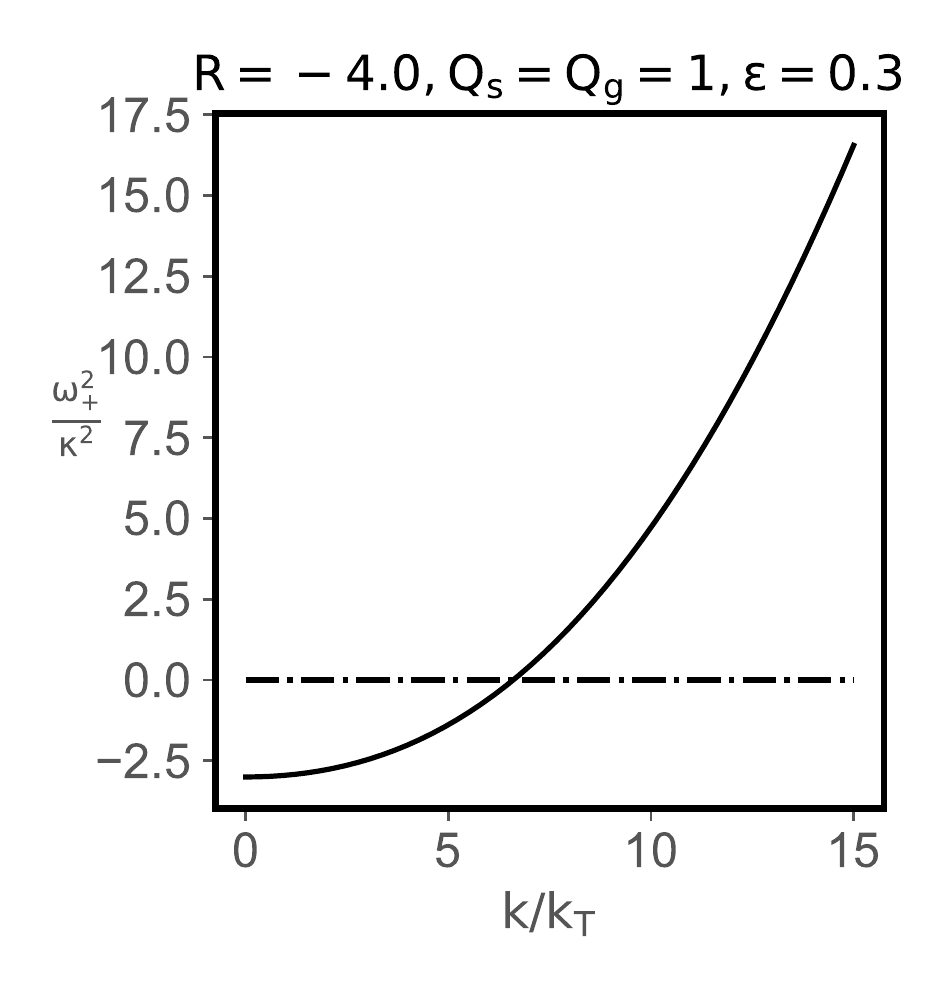}}
\end{tabular}
\end{center}
\caption{The above plots indicate the growth rate $\omega_{+}^{2}$ for varying value of compressive tidal field R= -0.5, -2.5, -4.0  value of 
$\epsilon$=0.3, 0.05 and $Q_{s}$=$Q_{g}$=1. }
\end{figure*}

From  Table 4 and Table 5 we can infer that a disruptive tidal fields tends to makes $\omega_{-(min)}$ more positive,thus making disc less prone to growth of instabilities, 
whereas a compressive tidal field tends to make $\omega_{-(min)}$ more negative and thus making the disc more susceptible to the growth of axis-symmetric perturbations. 
From table 4 we can see that growth rate $\omega_{-(min)}$ is positive for $Q_{s}=Q_{g}=2$, but  intense compressive tidal of order R= -2.5 at even very small gas fraction 
can make $\omega_{-(min)}$ negative. The growth rate of instabilities is enhanced immensely in case of galactic disc with high gas fraction aided by external compressive tidal
field. Finally from figure 12 it can be seen that under influence of intense compressive tidal fields even the $\omega_{+(min)}$ can become negative thus further enhancing 
the growth of instabilities in galactic disc.

\section{Conclusion}
In this work differential equations governing the growth of instabilities in two-fluid galactic disc under external tidal field have been derived and a criterion for 
appraising the stability of the two-fluid disc is presented. The modified stability criterion is applied to the understand the influence of the compressive and disruptive 
tidal fields, along with role of gas fraction on the stability of the galactic disc. We have also derived modified dispersion relation and studied the effect of the tidal fields 
on growth rate of instabilites in galactic disc, and interestingly found out that the $\omega_{+}^{2}$ can also become negative and can further destabilise the disc. 
Some possible applications of the results presented here are: understanding  the role of tidal fields in enhancing or quenching of star formation rates 
in interacting galaxies, understanding the role of dark matter haloes in stabilising gas rich low surface brightness galaxies,which without the tidal effects of the dark matter haloes will be unstable owing to large gas-fraction.

\small{\bibliographystyle{plainnat}}
\bibliography{example} 
\end{document}